\documentclass[runningheads]{llncs}
\usepackage{eccv}

\usepackage{eccvabbrv}
\usepackage{multirow}

\usepackage{graphicx}
\usepackage{booktabs}

\usepackage[accsupp]{axessibility}  %

\usepackage{hyperref}

\usepackage{orcidlink}

\usepackage{wrapfig}

\usepackage{docmute}

\begin{document}

\title{High-Quality Mesh Blendshape Generation from Face Videos via Neural Inverse Rendering} 

\titlerunning{High-Quality Mesh Blendshape Generation}

\author{Xin Ming\inst{1*}\orcidlink{0009-0007-9602-6078} \and
Jiawei Li\inst{2*}\orcidlink{0000-0001-7552-558X} \and
Jingwang Ling\inst{1}\orcidlink{0000-0001-8746-8578} \and
Libo Zhang\inst{1}\orcidlink{0009-0007-5291-9552} \and
Feng Xu\inst{1}\orcidlink{0000-0002-0953-1057}
}

\authorrunning{X. Ming et al.}

\institute{BNRist and School of Software, Tsinghua University \and
The Hong Kong University of Science and Technology
}

\maketitle

{\let\thefootnote\relax\footnotetext{* Both authors contributed equally to the paper}}

\begin{abstract}
Mesh-based facial blendshapes have been widely used in animation pipelines, while recent advancements in neural geometry and appearance representations have enabled high-quality inverse rendering. Building upon these observations, we introduce a novel technique that reconstructs mesh-based blendshape rigs from single or sparse multi-view videos, leveraging state-of-the-art neural inverse rendering. We begin by constructing a deformation representation that parameterizes vertex displacements into differential coordinates with tetrahedral connections, allowing for high-quality vertex deformation on high-resolution meshes. By constructing a set of semantic regulations in this representation, we achieve joint optimization of blendshapes and expression coefficients. Furthermore, to enable a user-friendly multi-view setup with unsynchronized cameras, we use a neural regressor to model time-varying motion parameters. 
Experiments demonstrate that, with the flexible input of single or sparse multi-view videos, we reconstruct personalized high-fidelity blendshapes. These blendshapes are both geometrically and semantically accurate, and they are compatible with industrial animation pipelines. Code and data are available at \url{https://github.com/grignarder/high-quality-blendshape-generation}.
\end{abstract}

\section{Introduction}
\label{sec:intro}

Synthesizing realistic 3D facial animations has long held significant applications in the movie and gaming industry. 
Accurate modeling of facial geometry and expression deformation constitutes a fundamental challenge for this task. 
In the industry, modeling usually involves a studio-level multi-view setup \cite{ghosh2011multiview,beeler2011high,debevec2000acquiring} to capture facial performances of real humans, along with the artist's manual effort to generate a facial rig. 
This facial rig is then imported into an animation pipeline \cite{DBLP:journals/mms/VilchisPMG23,DBLP:conf/svr/CruzT21} for game and movie production. 
VR and AR applications further require modeling facial rigs for a vast user base, necessitating an automated approach for facial modeling from widespread capture setups. 
One critical requirement is that the modeled face rig must be compatible with the animation pipeline to enable downstream animation applications.

\begin{figure}[htbp]
    \centering
    \includegraphics[width=\linewidth]{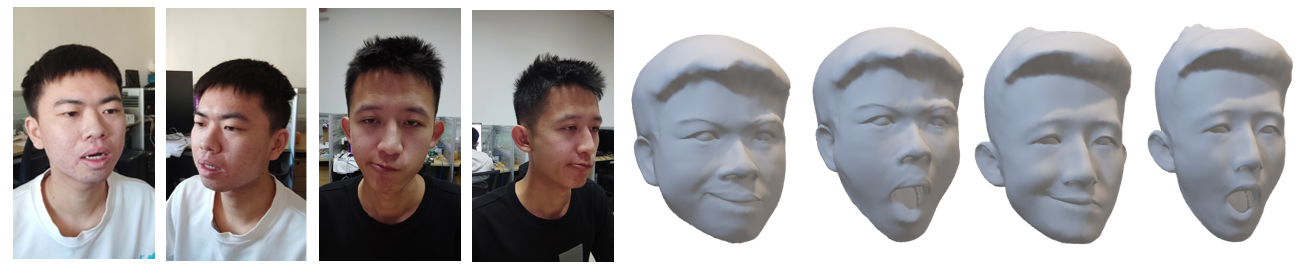}
    \caption{With the input of sparse multi-view face videos (shown on the left), our technique reconstructs personalized mesh-based blendshapes (examples shown on the right) that are ready to be used in the industrial animation pipeline.}
    \label{fig:teaser}
\end{figure}

RGB cameras are prevalent on everyday mobile devices, making them a popular choice for user-friendly facial reconstruction in numerous works. 
In addition to detecting facial landmarks from RGB inputs to fit facial statistical models \cite{DBLP:journals/tog/CaoHZ14,DBLP:conf/cvpr/ThiesZSTN16,DBLP:journals/tog/CaoWLZ13}, differentiable rendering \cite{DBLP:journals/corr/abs-2006-12057} can improve the reconstruction fidelity by harnessing dense pixel observations. 
However, overly-simplified rendering models cause under-fitting of facial material and arbitrary lighting, thereby negatively impacting the shape reconstruction quality. 
With the recent advancements in neural inverse rendering \cite{mildenhall2020nerf,DBLP:journals/cgf/TewariTMSTWLSML22}, techniques like neural facial avatars \cite{Gafni_2021_CVPR,INSTA:CVPR2023,grassal2021neural,zheng2022imavatar,Zheng2023pointavatar,DBLP:journals/corr/abs-2310-17519} can generate realistic animatable avatars from common RGB recordings.
However, these techniques do not rely on high-quality topology-consistent mesh representation and thus are not compatible with the industrial animation pipeline, impacting their practical utility.
To bridge the gap between realistic modeling and compatibility with current animation pipelines using easy recording setups, we, on one hand, represent dynamic facial modeling as a blendshape rig \cite{DBLP:conf/eurographics/LewisARZPD14}, consisting of topology-consistent facial meshes for various expressions. 
On the other hand, we optimize the blendshape with novel per-vertex deformation schemes to precisely match the generated animation to the facial performance in RGB videos (inverse rendering). 
Once converged, the obtained blendshape can be imported into animation software (e.g. Blender \cite{Blender}) to generate realistic person-specific facial animations for industrial applications.

To achieve high-quality shape reconstruction and animation by optimizing the blendshape rig via neural inverse rendering, we propose techniques to solve three unaddressed issues. 
The first arises in optimizing per-vertex deformations of a high-resolution mesh, which can be non-smooth and suffer from self-intersections. 
By applying differential coordinates to parameterize blendshape meshes augmented with tetrahedral connections, we facilitate gradient propagation along topologically and spatially adjacent vertices, ensuring smooth deformation.
Secondly, there is an ambiguity in optimizing either expression bases or coefficients to fit users' arbitrary facial performance, and prior methods \cite{grassal2021neural,Gafni_2021_CVPR,INSTA:CVPR2023,zheng2022imavatar,DBLP:journals/corr/abs-2310-17519} typically circumvent this by excluding expression coefficients from the optimization (estimating them through a pre-processing step \cite{DBLP:conf/cvpr/ThiesZSTN16}) thus only reaching local optima. 
We aim to improve convergence by joint optimization with novel regularization techniques that enforce the symmetry, sparsity, and semantics of expression bases to solve the ambiguity.
Thirdly, multi-view inputs are useful for accurately reconstructing non-rigid facial deformations \cite{DBLP:conf/nips/GaoLTRK22}, but previous research usually does not presume that multi-view inputs are readily available, as they are typically linked with complex procedures such as synchronization and color correction. 
We incorporate sparse multi-view inputs from unsynchronized smartphones by utilizing a neural regressor to model time-dependent motion parameters, implicitly ensuring temporal synchronization.
In summary, our contributions include: 
\begin{itemize}
    \item A video-based facial rigging technique that bridges traditional animation pipelines and neural inverse rendering to achieve high-quality animation-ready facial rig reconstruction from single or sparse multi-view videos (as shown in \cref{fig:teaser}), and
    \item a novel blendshape deformation technique that parameterizes differential coordinates augmented with tetrahedral connections, involving a set of semantic regularization into a joint optimization.
\end{itemize}

\section{Related Work}
\label{relatedWork}

\begin{figure*}[tp]
    \centering
    \includegraphics[width=\textwidth]{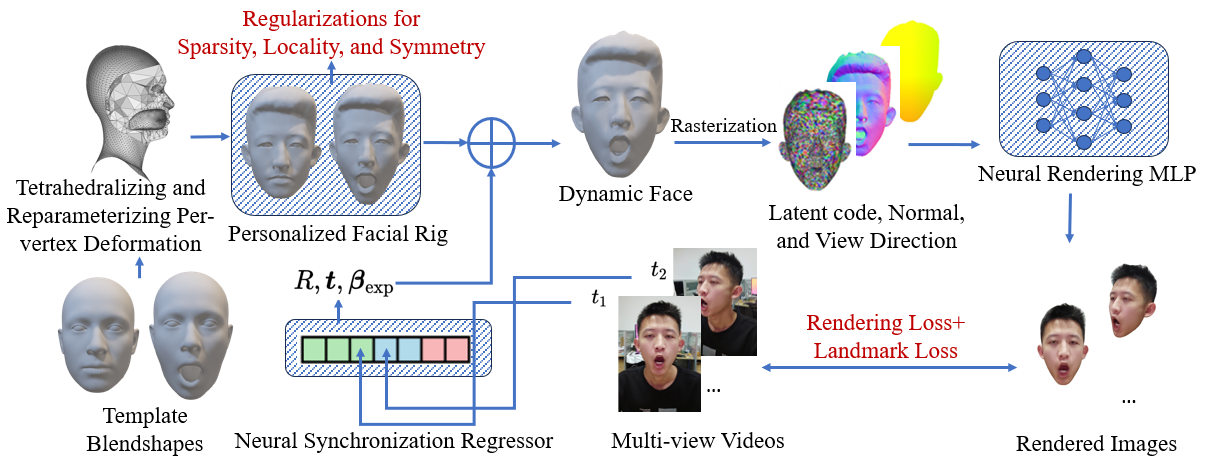}
    \caption{Method pipeline. We model the human head as a person-specific facial rig that includes a neutral face and a set of blendshapes. This rig is derived from template blendshapes through tetrahedralizing and reparameterizing per-vertex deformation. The head poses $R$,$\boldsymbol{t}$ and expression coefficients $\boldsymbol{\beta}_{exp}$ are regressed from the timestamps corresponding to each frame by a neural synchronization regressor, which achieves implicit synchronization between the multi-view, not fully synchronized videos. Combined with the facial rig, the dynamic face geometry is obtained. Afterwards, a neural rendering MLP renders the corresponding images according to the latent codes, normals, and view directions acquired through differentiable rasterization. Finally, we leverage the rendering loss, landmark loss, and rigging regularization terms to jointly optimize the facial rig, the neural regressor, and the neural rendering MLP.}
    \label{fig:pipeline}
\end{figure*}

{\bf 3D Facial Performance Capture.}
Many studies have been devoted to generating realistic 3D animations from users' facial performance.
High-quality facial animation can be reconstructed through a studio-level multi-view setup \cite{beeler2011high,DBLP:journals/tog/BradleyHPS10,DBLP:journals/cgf/FyffeNHSBJLD17}. However, this involves intricate procedures for dozens of professional cameras, including synchronization and color correction.
To enable facial performance capture using ubiquitous devices, 
morphable models \cite{DBLP:journals/tog/VlasicBPP05,DBLP:conf/siggraph/BlanzV99} are fitted from monocular RGB or RGBD videos \cite{DBLP:journals/tog/CaoHZ14,DBLP:journals/tog/CaoWLZ13,DBLP:journals/tog/WeiseBLP11}.
To achieve a more personalized facial geometry beyond the morphable model, fine-level displacements are introduced on the facial mesh to synthesize nuanced facial details \cite{DBLP:journals/tog/GarridoVWT13,DBLP:journals/tog/CaoBZB15,DBLP:conf/si3d/MaD19}.
Due to the inherent ambiguity in non-rigid facial reconstruction from monocular input, deformation is highly constrained. 
In attempts to address such limitation, efforts extend to sparse views and observe reconstruction improvements \cite{DBLP:journals/tog/CaoATCSSS21,DBLP:journals/tog/ValgaertsWBST12}. 
Establishing a simple and user-friendly sparse view setup remains an active research topic.
While the aforementioned approaches can reconstruct dynamic facial geometries, additional efforts are required to organize the performance data into a facial rig for convenient editing and synthesis of novel facial animations.

{\bf 3D Facial Rigging.}
Rigging aims to generate a personalized facial expression model from the user's performance, typically represented using blendshapes \cite{DBLP:conf/eurographics/LewisARZPD14} for compatibility with the animation pipeline.
Deformation transfer \cite{sumner2004deformation} can personalize template blendshapes from a neutral expression mesh. 
Furthermore, data-driven priors are utilized to predict personalized expression bases from a neutral scan or image \cite{DBLP:journals/tog/LiKZHB020,DBLP:conf/cvpr/Yang0WHSYC20} .
To achieve higher degrees of personalization from more observations, some works \cite{li2010example} take input from multiple scans with predefined expressions, while some \cite{ichim2015dynamic} require users to make specific key expressions during the capture process.
Efforts on exploring more user-friendly capture procedures \cite{garrido2016reconstruction,DBLP:journals/tog/BouazizWP13,DBLP:journals/tog/LiYYB13,wang2020emotion} focus on utilizing performance sequences where users make arbitrary facial expressions to generate expression blendshapes.
Updating blendshapes requires careful design to avoid mesh non-smoothness, thus techniques such as reduced subspace \cite{DBLP:journals/tog/BouazizWP13} and corrective shapes \cite{DBLP:journals/tog/LiYYB13,garrido2016reconstruction} are employed to constrain deformations.
To resolve the ambiguity between expression bases and expression coefficients, semantic emotion priors are proposed to constrain expressions \cite{wang2020emotion}.
Some deep-learning-based methods \cite{chaudhuri2020personalized,tewari2019fml} propose an end-to-end framework
 that learns a personalized face model from a corpus of in-the-wild videos.
Our work introduces a vertex deformation representation that enables high-fidelity deformation of blendshapes while enforcing smoothness. We also design constraints to maintain semantic coherence in expression blendshapes.

{\bf Neural Inverse Rendering.}
Differentiable rendering \cite{DBLP:conf/iccv/Liu0LL19,DBLP:journals/corr/abs-2007-08501,DBLP:journals/tog/LaineHKSLA20} can leverage gradient backpropagation to optimize geometry, material and lighting to achieve inverse rendering-based reconstruction.
Facial materials, influenced by subsurface scattering \cite{DBLP:conf/rt/DonnerJ06}, are challenging to represent using simplified rendering models, which can lead to underfitting in differentiable rendering.
Recent advances in neural rendering \cite{DBLP:journals/tog/ThiesZN19,mildenhall2020nerf,DBLP:conf/cvpr/WorchelDHSFE22} bypass this limitation by directly modeling the emitting radiance via neural networks, achieving realistic novel-view synthesis \cite{DBLP:journals/tog/ThiesZN19,mildenhall2020nerf} and reconstruction \cite{DBLP:conf/cvpr/WorchelDHSFE22} of static objects.
Dynamic object modeling is achieved via neural deformation fields \cite{DBLP:conf/nips/CaiFFWZ22,DBLP:conf/cvpr/ShaoZTL0L23}, but do not incorporate expression-driven retargeting.
Some works \cite{Gafni_2021_CVPR,INSTA:CVPR2023,DBLP:conf/cvpr/AtharXSSS22,DBLP:conf/siggraph/XuWZ0L23,DBLP:journals/tog/GaoZXHGZ22,DBLP:conf/cvpr/ChenOB023} extend NeRF \cite{mildenhall2020nerf} to expression-driven dynamic faces. 
However, the density-based representation employed by their methods lacks explicit geometric regularization, often lowering the quality of novel views.
Alternative representations such as implicit fields, point clouds and 3D Gaussians \cite{zheng2022imavatar,Zheng2023pointavatar,xiang2024flashavatar,qian2024gaussianavatars} are explored, but compatibility with animation pipelines remains a challenge.
Neural Head Avatars \cite{grassal2021neural} can obtain a mesh representation after a long training time, but with a primary focus on rendering quality rather than accurate geometry reconstruction, frequently leading to details being baked in textures.
FLARE \cite{DBLP:journals/corr/abs-2310-17519} explores the use of mesh-based representation for fast learning of facial avatars.
Compared to their works, our emphasis lies more on the geometric quality and compatibility of animation. Therefore, we jointly optimize blendshapes and expression coefficients, incorporate regularization to maintain the semantics of the updated blendshapes, and achieve accurate reconstruction evaluated by point-to-plane distance.

\section{Method}
We aim to reconstruct personalized mesh-based blendshapes from RGB videos.
Personalization involves per-vertex deformation applied to the blendshapes. 
We propose a deformation representation, outlined in \cref{sec:deformable_rig}, to ensure smoothness and prevent self-intersection for high-resolution meshes.
Based on the representation, the deformations are further regularized, introduced in \cref{sec:rigging_reg}, to maintain the semantics of the expression blendshapes.
The sparse multi-view inputs, which are used to guide the deformations, are implicitly synchronized by a neural synchronization regressor illustrated in \cref{sec:sparse_multi_view}. 
Additionally, with a neural rendering pipeline in \cref{sec:neural_rendering} to render the animated faces, we compare the rendered faces with the input to reconstruct the blendshape deformation. 
To be specific, the reconstruction is solved by the joint optimization of the blendshape deformation, the rendering network, and the synchronization regressor in \cref{sec:joint_opt}.
\cref{fig:pipeline} represents an overview of our method.

\subsection{Vertex Deformations with Tetrahedral Connections} %
\label{sec:deformable_rig}
The facial shape is represented by a mesh where a neutral face $\mathbf{b}_n$
describes its identity and blendshapes \cite{DBLP:conf/eurographics/LewisARZPD14} describe its expression deformation. 
This blendshape model represents a face with a specific expression as $\mathbf{b}_\beta=\mathbf{b}_n+B_{\exp } \boldsymbol{\beta_{\text{exp}}}$, where $\mathbf{b}_n$ denotes the user-specific neutral face, $B_{\exp } \in$ $\mathbb{R}^{3 N \times M_{\exp }}$ denotes the blendshape model, and $\boldsymbol{\beta}_{\text{exp}} \in \mathbb{R}^{M_{\text {exp }}}$ represents the expression coefficients. 
Our objective is to generate a person-specific facial rig consisting of a neutral face $\mathbf{b}_{n}^{\ast}$ and a set of blendshapes $B_{\text{exp}}^{\ast}$ by solving per-vertex deformation applied to $\mathbf{b}_{n}$ and $B_{\text{exp}}$ from a base blendshape model (ICT Face Model \cite{ICTli2020learning} in our experiments).

However, directly optimizing per-vertex deformation poses challenges for convergence \cite{Nicolet2021Large}.
To ensure smoothness and prevent self-intersection cavities, we devise a vertex parameterization that implicitly satisfies volumetric Laplacian regularization.
First, we parameterize vertex displacements into differential coordinates \cite{DBLP:conf/sgp/SorkineCLARS04}, inspired by \cite{Nicolet2021Large}.
The parameterization propagates vertex gradients to neighboring vertices based on mesh connectivity, effectively enforcing smooth deformation.
However, there is no gradient propagation between spatially adjacent but not directly connected vertices, and mesh self-intersection can still occur.
Therefore, we augment mesh connectivity via internal tetrahedral filling.
Specifically, we use TetGen \cite{tetgen} to fill the closed space between the surface and corresponding internal sockets with tetrahedras, preventing interpenetration due to large deformations.
More details about tetrahedral filling can be found in our supplementary document.
We use $\Phi$ to denote the process of tetrahedralizing and reparameterizing per-vertex deformation. The personalized neutral face is represented as $\mathbf{b}_{n}^{\ast}=\Phi \left( \mathbf{b}_n \right)$.
Blendshapes are deformed similarly as $B_{\text{exp} }^{\ast }=\Phi \left( B_{\text{exp} }\right)$.

\textit{Discussion.}
\cite{grassal2021neural, Zheng2023pointavatar} employ MLPs to regress deformations from canonical vertex coordinates, observing that the output deformations exhibit spatial smoothness.
We attribute this phenomenon to the shared network among vertices, where during backpropagation, the gradient of one vertex influences others, with a greater impact on adjacent vertices \cite{DBLP:conf/nips/TancikSMFRSRBN20}.
We have employed a network-free method that achieves similar effects, propagating vertex gradients to topologically and spatially adjacent vertices. This approach is memory-efficient, faster, and suitable for applications with multiple ($M_{\text {exp }}=53$) blendshape bases.

\subsection{Rigging Regularization}
\label{sec:rigging_reg}
Blendshapes have clear semantics due to their connection with facial action units \cite{semantic_blendshape}. However, the semantics may be corrupted due to ambiguity as we optimize both expression coefficients $\boldsymbol{\beta}_{\text{exp}}$ and blendshapes $B_{\text{exp} }^{\ast }$ simultaneously. To this end, we propose regularizations based on three principles, namely locality, sparsity, and symmetry, to ensure that we obtain a semantically consistent rig.

\textbf{Locality.} 
Each blendshape corresponds to an action unit, and its deformation has a localized influence region.
Inspired by \cite{chaudhuri2020personalized}, the update of a blendshape should be concentrated on its original activation region. 
To this end, we first compute the per-vertex deforming weights $W \in \mathbb{R}^{3 N \times M_{\exp }}$ based on the initial blendshapes given by
\begin{equation}
W\left(3i:3i+2,j\right) = \exp \left( -\dfrac{\left\| B_{\exp} \left( 3i:3i+2,j\right) \right\| _{2}}{a}\right)
\end{equation}
where $a$ is a hyperparameter controlling the smoothness of the activation region boundary.

The weight is used to compute the locality loss defined as
\begin{equation}
\mathcal{L}_{\text{locality}}=\left\|  W\odot \left(  B_{\text{exp} }^{\ast }-B_{\exp } \right) \right\| _{F}
\end{equation}
where $\odot$ denotes element-wise multiplication.

\textbf{Sparsity.} The dynamic facial deformation should be explained by only a few blendshapes. 
When multiple blendshape coefficients are wrongly activated during optimization, a sparsity regularization on blendshapes can prevent the deformation to be averaged into multiple blendshapes.
The sparsity loss is defined as:
\begin{equation}
\mathcal{L}_{\text{sparsity}}=\left\| B_{\text{exp}}^{\ast }-B_{\text{exp} }\right\| _{p}
\end{equation}
with $p<1$. We use $p=0.75$ in the experiments.

\textbf{Symmetry.} The blendshapes which are symmetric for the left and right faces should still maintain symmetry. We manually select the symmetric ones from the initial blendshapes, and only update their left half faces. The right half faces are obtained by symmetry.

\subsection{Sparse Multi-View Handling}
\label{sec:sparse_multi_view}
Accurate modeling of dynamic faces from monocular videos is an ill-posed problem \cite{DBLP:conf/nips/GaoLTRK22}. 
However, increasing the number of viewpoints often incurs cumbersome setups such as synchronization.
Conversely, we allow unsynchronized RGB videos captured from mobile phones as input.
To address the issue of incomplete time synchronization among multiple devices, we propose to use a one-dimensional Instant-NGP\cite{mueller2022instant} to store temporal information to implicitly ensure synchronization.
Specifically, for each viewpoint $k$, we record the video start time $t_{s}^{k}$ from the system clock of the mobile phone. 
The time of the $i$th frame can be calculated as $t_{i}^{k} = t_{s}^{k} + \dfrac{i}{r_{k}}$, where $r_k$ is the frame rate. 
$t_{i}^{k}$ will be used to regress parameters containing face rotation $R_i^k$, translation $\boldsymbol{t}_i^k$, and expression coefficients $\boldsymbol{\beta}_i^k$ with the neural regressor as:
\begin{equation}
R_i^k,\boldsymbol{t}_i^k,\boldsymbol{\beta}_i^k = \text{Grid}\left( t_i^k\right)
\end{equation}
Compared to another viewpoint $k'$, while $t_i^k$ and $t_i^{k'}$ are not captured at the same time, they have independent motion parameters, and the neural regressor ensures smoothness for temporally close parameters.

To address the exposure difference among different viewpoints, we assign a learnable latent code for each camera $\boldsymbol{h}_k$ when rendering. Details will be explained in the next section.

\subsection{Mesh-based Neural Deferred Rendering}
\label{sec:neural_rendering}
Mesh-based face models enable us to perform efficient rendering using differentiable rasterization \cite{DBLP:journals/tog/LaineHKSLA20}.
However, overly simplified rendering models may suffer from underfitting due to the complex material of the face and arbitrary lighting. Motivated by \cite{worchel:2022:nds}, we use a technique that combines neural rendering and deferred rendering from real-time rendering pipelines. Specifically, a latent code is assigned to each mesh vertex, which represents the neural texture. In the rendering process, the mesh is first rasterized, yielding the triangle indices and barycentric coordinates for each pixel, which are used to interpolate the latent codes, vertex normals and view directions. Then, we use a learnable MLP-based shader to regress the per-pixel RGB color:
\begin{equation}
    f_{\boldsymbol{\theta} }\left( \boldsymbol{z},\boldsymbol{n},\boldsymbol{\omega},\boldsymbol{h}_k\right) \in \left[ 0,1\right] ^{3}
\end{equation}
where $\boldsymbol{z}$ denotes the latent code, $\boldsymbol{n}$ denotes the normal, $\boldsymbol{\omega}$ denotes the view direction, $\boldsymbol{h}_k$ denotes the learnable latent code assigned to the $k$-th viewpoint and $\boldsymbol{\theta}$ denotes the network parameters.

\subsection{Joint Optimization}
\label{sec:joint_opt}
Our optimization objective integrates multiple loss components to collectively optimize all trainable parameters from randomly initialized values, including the facial rig, the neural regressor, and the neural shader. The formulation of the joint optimization objective is expressed as follows:
\begin{equation}
\begin{aligned} \mathcal{L}_{\text {total }}= & \mathcal{L}_{\text {ldmk }}+\mathcal{L}_{\text {mask }}+\mathcal{L}_{\text {photometric }}\\ & +\mathcal{L}_{\text {Laplacian }} +\mathcal{L}_{\text {locality}}\\ & +\mathcal{L}_{\text {sparsity}} +\mathcal{R}_{\text {exp }} + \mathcal{R}_{\text {neutral}}\end{aligned}
\end{equation}
This objective function encapsulates various aspects, including landmark loss $\mathcal{L}_{\text {ldmk }}$, mask loss $\mathcal{L}_{\text {mask }}$, photometric loss $\mathcal{L}_{\text {photometric }}$, Laplacian loss $\mathcal{L}_{\text {Laplacian }}$, expression regularization $\mathcal{R}_{\text {exp }}$, and regularization for template deformation $\mathcal{R}_{\text {neutral}}$. $\mathcal{L}_{\text {sparsity}}$ and $\mathcal{L}_{\text {locality}}$have been explained in the previous section. $\mathcal{L}_{\text {ldmk }}$ enforces accurate prediction of facial landmarks, 
\begin{equation}
\mathcal{L}_{\text {ldmk }} = \frac{1}{N} \sum_{i=1}^{N} \| \hat{v}_i - v_i \|_1
\end{equation}
where $\hat{v}$  indicates the landmarks projected on images and $v$ indicates the detected $N$ landmarks. 
$\mathcal{R}_{\text {exp }}$ serves as the sparsity regularizer, 
\begin{equation}
\mathcal{R}_{\text {exp }}=\left\|\boldsymbol{\beta}_{\text{exp}}\right\|_1
\end{equation}
where $\boldsymbol{\beta}_{\text{exp}}$ is the expression coefficient.
$\mathcal{L}_{\text {mask }}$ ensures the alignment of rendered masks $\hat{M}$ and segmented masks $M$, and $\mathcal{L}_{\text {photometric }}$ enforces consistency between rendered images $\hat{I}$ and captured images $I$
\begin{equation}
\mathcal{L}_{\text {mask }}=\left\|\hat{M}-M\right\|_{1}
\end{equation}
\begin{equation}
\mathcal{L}_{\text {photometric }}=\left\|M \odot\left(\hat{I}-I\right)\right\|_{1}
\end{equation}
where $\odot$ denotes element-wise multiplication.
$\mathcal{L}_{\text {Laplacian }}$ enforces smoothness of latent codes between adjacent vertices.
\begin{equation}
\mathcal{L}_{\text {Laplacian }} =\left\| LU\right\| ^{2}
\end{equation}
where $L$ is the Laplacian matrix and $U$ denotes the per-vertex latent codes, with its $i$-th row storing the latent code of the $i$-th vertex. $\mathcal{R}_{\text {neutral}}$ constrains deformation of the neutral face. 
\begin{equation}
\mathcal{R}_{\text {neutral}} = \left\| \mathbf{b}_n^{\ast }-\mathbf{b}_n\right\| _{2}^{2}
\end{equation}

This comprehensive optimization objective facilitates the joint refinement of our pipeline. In our experiment, the $\mathcal{L}_{\text {ldmk }}$ (including the landmarks on eye balls) and $\mathcal{R}_{\text {exp }}$ are initially activated to obtain a coarse alignment. After a number of epochs, we proceed to enable all the loss components.

\section{Experiments}

In this section, we first describe the implementation details of our method and provide information about the used datasets. 
Next, we qualitatively and quantitatively compare the accuracy of geometric reconstruction with previous works. 
We then conduct ablation studies to assess the impact of the deformation representation on vertex optimization and the role of semantic regularization in constraining expression bases. 
Finally, we demonstrate the application of our method in animation, including expression retargeting and novel-view synthesis.
More results can be found in our supplementary document and video.
\subsection{Implementation Details}
For the input videos, we use Facer \cite{zheng2022farl} to obtain the facial landmarks and masks. We use a three-layer MLP as the neural renderer, which has 64 hidden units and uses ReLU as the activation function. In the hierarchical grids of our neural synchronization regressor, we use 6 grid scales with a base resolution of 8,
and we use 4 channels per level. We use Nvdiffrast \cite{DBLP:journals/tog/LaineHKSLA20} as the differentiable rasterizer. For the neural renderer and the regressor, we use an Adam \cite{kingma2014adam} optimizer with $\eta = 1e^{-3}$ and $\beta = \left( 0.9,0.999\right)$. The facial rig is updated using an AdamUniform \cite{Nicolet2021Large} optimizer, with the same parameters as the Adam optimizer. We train our model for 200 epochs, with all loss functions activated for the last 120 epochs.

\subsection{Datasets and Metrics}
\textbf{Datasets} 
We capture our dataset using four mobile phones for qualitative comparisons. 
Additionally, we conduct qualitative and quantitative evaluations on the Multiface \cite{wuu2022multiface} and NeRSemble \cite{kirschstein2023nersemble} datasets, which feature high-quality multi-view captures of different identities with rich expressions. 
We utilize Meta\-Shape \cite{metashape2023} to reconstruct accurate 3D scans from all available views of the two datasets (38 in \cite{wuu2022multiface} and 16 in \cite{kirschstein2023nersemble}) as the ground truth. 
For each dataset, we manually select four views as the inputs to simulate the sparse-view setup, like \cite{DBLP:conf/cvpr/ShaoZTL0L23}.
All experiments in the main paper are conducted with four-view inputs. 
We present the experimental results under a single view in the supplementary materials to demonstrate that our method is also applicable for easier setup.

\textbf{Evaluation Metrics}
We adopt the evaluation metrics from \cite{wu2019mvf} to compute point-to-plane L2 errors at facial regions between reconstructed 3D shapes and ground-truth 3D scans.
We report reconstruction errors averaged across all frames in a video sequence.

\begin{table}[t]
\centering
\begin{tabular}{lllll}
\hline
\multirow{2}{*}{point-to-plane error(mm)} & \multicolumn{2}{l}{Multiface} & \multicolumn{2}{l}{NeRSemble} \\ \cline{2-5} 
                                          & Mean          & Std           & Mean          & Std           \\ \hline
NHA                                       & 3.76          & 0.13          & 4.98          & 0.36          \\
PointAvatar                               & 7.66          & 0.28          & 7.22          & 0.34          \\ 
FLARE                                     & 5.61          & 0.21          & 5.88          & 0.33          \\
HRN                                       & 4.37          & 0.14          & 4.53          & \textbf{0.19}          \\
Ours                                      & \textbf{2.31} & \textbf{0.05} & \textbf{2.73} & 0.26 \\ \hline
\end{tabular}
\caption{Quantitative comparison in point-to-plane errors among NHA, PointAvatar, FLARE, HRN and our method on the NeRSemble and MultiFace datasets.}
\label{table:quantitative_comparison}
\end{table}

\subsection{Comparisons}
\begin{figure}[tp]
    \centering
    \includegraphics[width=\linewidth]{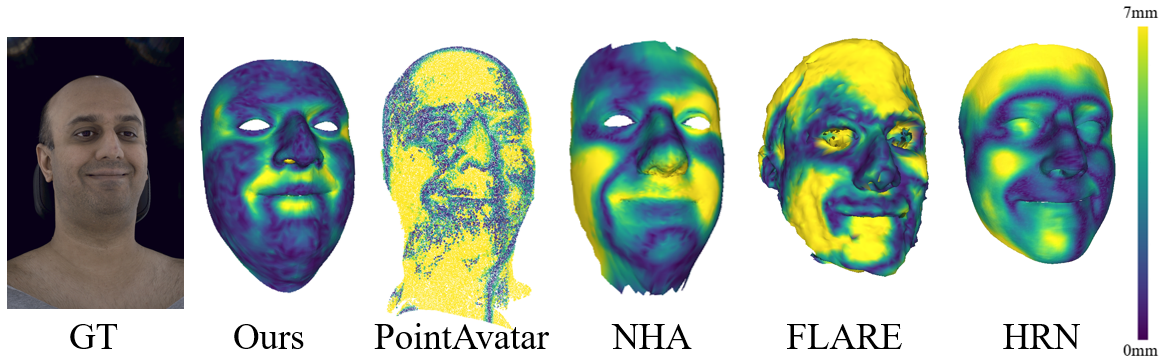}
    \caption{Visualization of the point-to-plane error heatmaps for PointAvatar, NHA, FLARE, HRN, and our method.}
    \label{fig:error_heatmap}
\end{figure}

\begin{figure}
    \centering
    \includegraphics[width=0.8\linewidth]{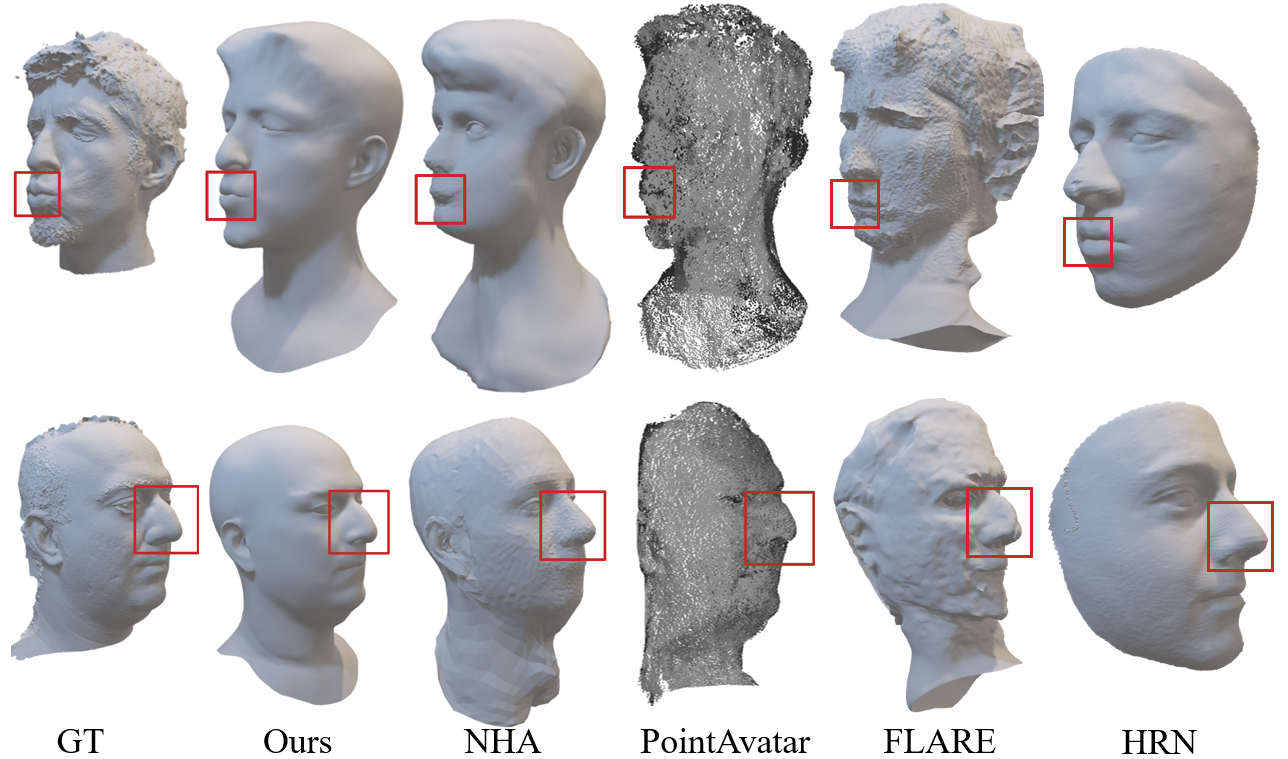}
    \caption{Comparisons of identity and expression-related facial details between our method and other baselines. %
    }
    \label{fig:recon_details}
\end{figure}

To evaluate the accuracy of the geometric reconstruction, we perform qualitative and quantitative comparisons on the reconstruction results using the Multiface \cite{wuu2022multiface} and NeRSemble \cite{kirschstein2023nersemble} datasets.
We choose to compare NHA \cite{grassal2021neural}, PointAvatar \cite{Zheng2023pointavatar} and FLARE \cite{DBLP:journals/corr/abs-2310-17519} as they represent the latest works on face avatars based on explicit shape representation. PointAvatar \cite{Zheng2023pointavatar} claims to achieve comparable geometry reconstruction with \cite{zheng2022imavatar}. Works such as \cite{Gafni_2021_CVPR,INSTA:CVPR2023,DBLP:conf/cvpr/AtharXSSS22,DBLP:conf/siggraph/XuWZ0L23,DBLP:journals/tog/GaoZXHGZ22} achieve high-quality rendering, but their density-based representations are not suitable for direct comparisons. We also compare our method with HRN \cite{lei2023hierarchical}, which is trained on large-scale in-the-wild images for accurate face reconstruction and can accept multi-view image inputs.
We modified the baselines so that all methods use input from four views.
As shown in Table \ref{table:quantitative_comparison}, our method surpasses other methods in point-to-plane errors on both datasets.
Lower errors are also evident in the visualized heatmaps in \cref{fig:error_heatmap}, where we achieve more accurate reconstruction, especially in the forehead and nose regions.
In \cref{fig:recon_details}, a qualitative comparison of the reconstruction results for identity and expression-specific facial details is presented.
In the first row, our method reconstructs a more personalized puckering expression.
In the second row, our method successfully reconstructs the aquiline nose, which is a distinctive geometric feature specific to the input identity.

\begin{figure}
    \centering
    \begin{subfigure}[b]{0.95\linewidth}
        \includegraphics[width=\linewidth]{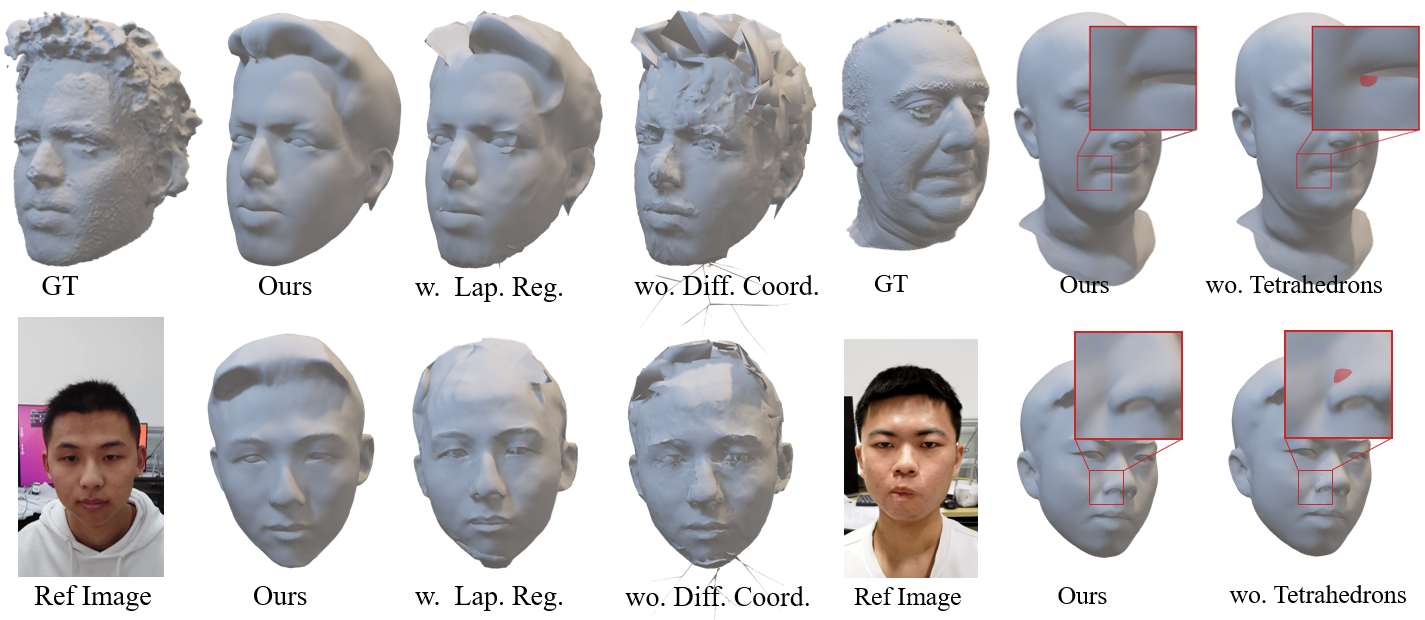}
    \end{subfigure}
    \caption{Evaluating the effectiveness of the blendshape deformation representation, including differential coordinate reparameterization and tetrahedral connections.}
    \label{fig:ablation_deformation}
\end{figure}

\begin{figure*}[tp]
    \centering
    \includegraphics[width=\linewidth]{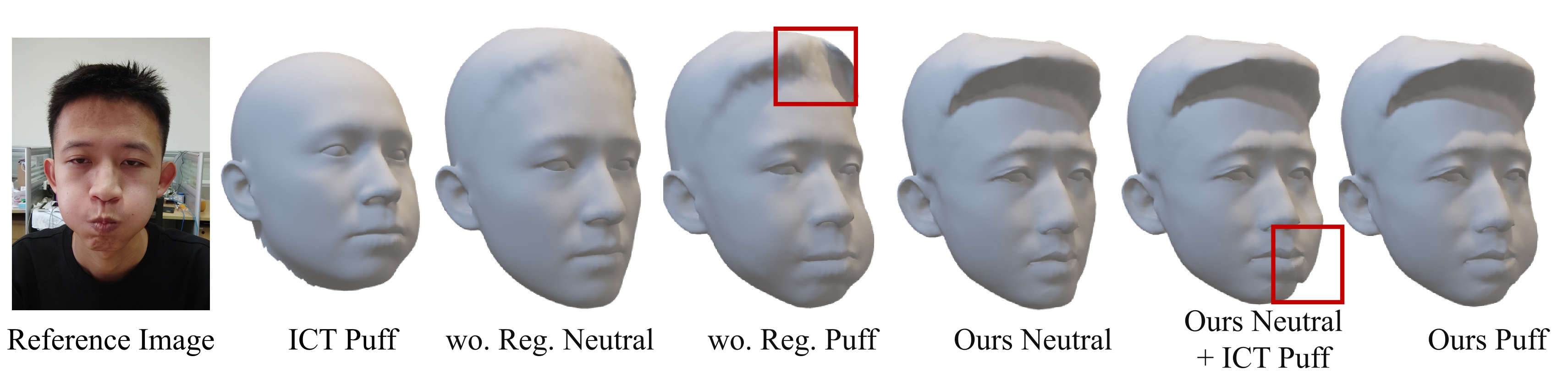}
    \caption{Blendshapes of neutral and cheek puffing expressions obtained by different solutions. The results reveal that our method not only correctly encodes the identity information in the neutral blendshape but also encodes the single-sided puffing expression in its corresponding blendshape.}
    \label{fig:rigging_result}
\end{figure*}

\begin{figure}[tp]
    \centering
    \includegraphics[width=0.7\linewidth]{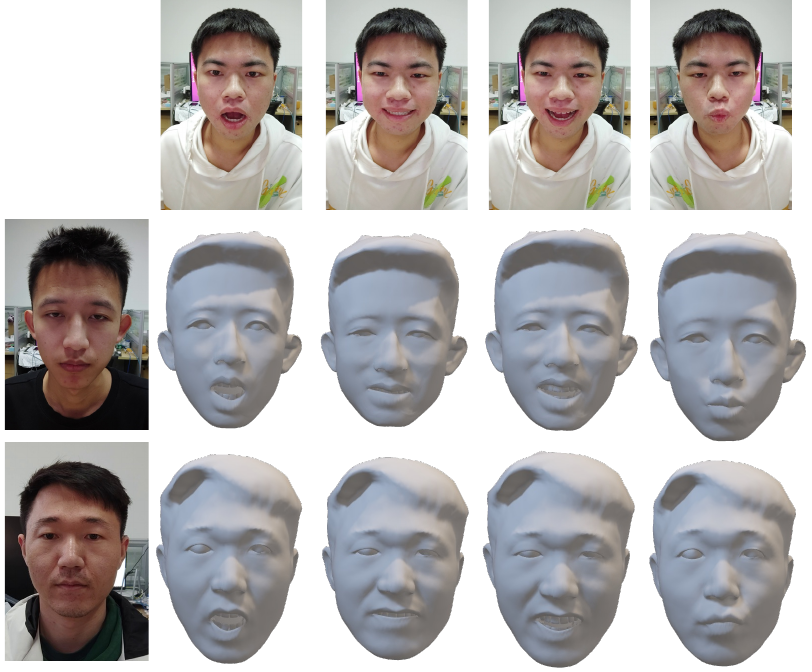}
    \caption{Retargeting results of our personalized facial rig. The first row shows the source expressions. The following rows show the retargeting results, where the images in the first column show the neutral expression of the target identities.}
    \label{fig:retargeting}
\end{figure}

\subsection{Ablation Study}
To test the necessity of the blendshape deformation representation in preserving the mesh's desirable properties, we present geometric reconstruction results under different settings.
We compare the reconstruction results of our method with: (1) without using differential coordinates and (2) with tetrahedral connections disabled. The results are then compared against the full pipeline.
As shown in the left half of \cref{fig:ablation_deformation}, the utilization of differential coordinates in the optimization process significantly enhances the smoothness of the face surface, effectively eliminating numerous artifacts while preserving geometry accuracy.
Replacing differential coordinates with a Laplacian regularizer could also increase smoothness, but it fails to prevent self-intersections and geometrically mismatches the target (third column).
The right half in \cref{fig:ablation_deformation} illustrates the results of using tetrahedral connections during the vertex deformation process.
When the user exhibits extreme facial expressions, such as a puckered mouth, there is a risk of penetration between the mouth socket and the facial surface, especially for high-resolution meshes. 
The twisting of the nose, due to the presence of the nasal cavity, may result in similar issues.
Even if it occurs in a limited region, it poses significant challenges for artists in refining and adjusting the reconstructed facial rigs.
By establishing tetrahedral connections between surface points and internal socket points, we effectively mitigated the penetration without compromising the accuracy of deformation.

To evaluate the impact of blendshape updates and semantic regularization in the updates, we visualize the obtained expression bases under different settings in \cref{fig:rigging_result}.
The first column showcases a frame from the input sequence where the user makes a puffy expression. Our objective is to update the personalized one-sided puffy expression basis based on the inputs.
If expressions are made only in the expression space of the ICT morphable model (second column), the resulting face deviates significantly from the user's identity, lacking personalization.
If the expression basis is updated without applying semantic regularization, identity-specific hair details are missing on the neutral face (third column). The relevant detail components appear inappropriately in the expression basis (fourth column, highlighted by red boxes).
When applying semantic regularization, our method can reconstruct a high-quality neutral face (fifth column) that includes all identity-related facial details.
However, if the expression basis is not updated and template blendshapes are directly applied to deform the personalized neutral face, artifacts due to mismatched deformations occur in the deformed region (sixth column, highlighted by red boxes).
After updating the expression basis and applying semantic regularization, our method synthesizes high-quality personalized expression bases (seventh column).

\subsection{Applications}
In this section, we showcase the animation applications of the reconstructed results, including expression retargeting and novel-view synthesis.

{\bf Expression retargeting.}
The reconstructed geometry, represented by blendshapes with consistent topology, adheres to the format of animation pipelines. 
\begin{wrapfigure}{l}{0.28\linewidth}
    \centering
    \includegraphics[width=\linewidth]{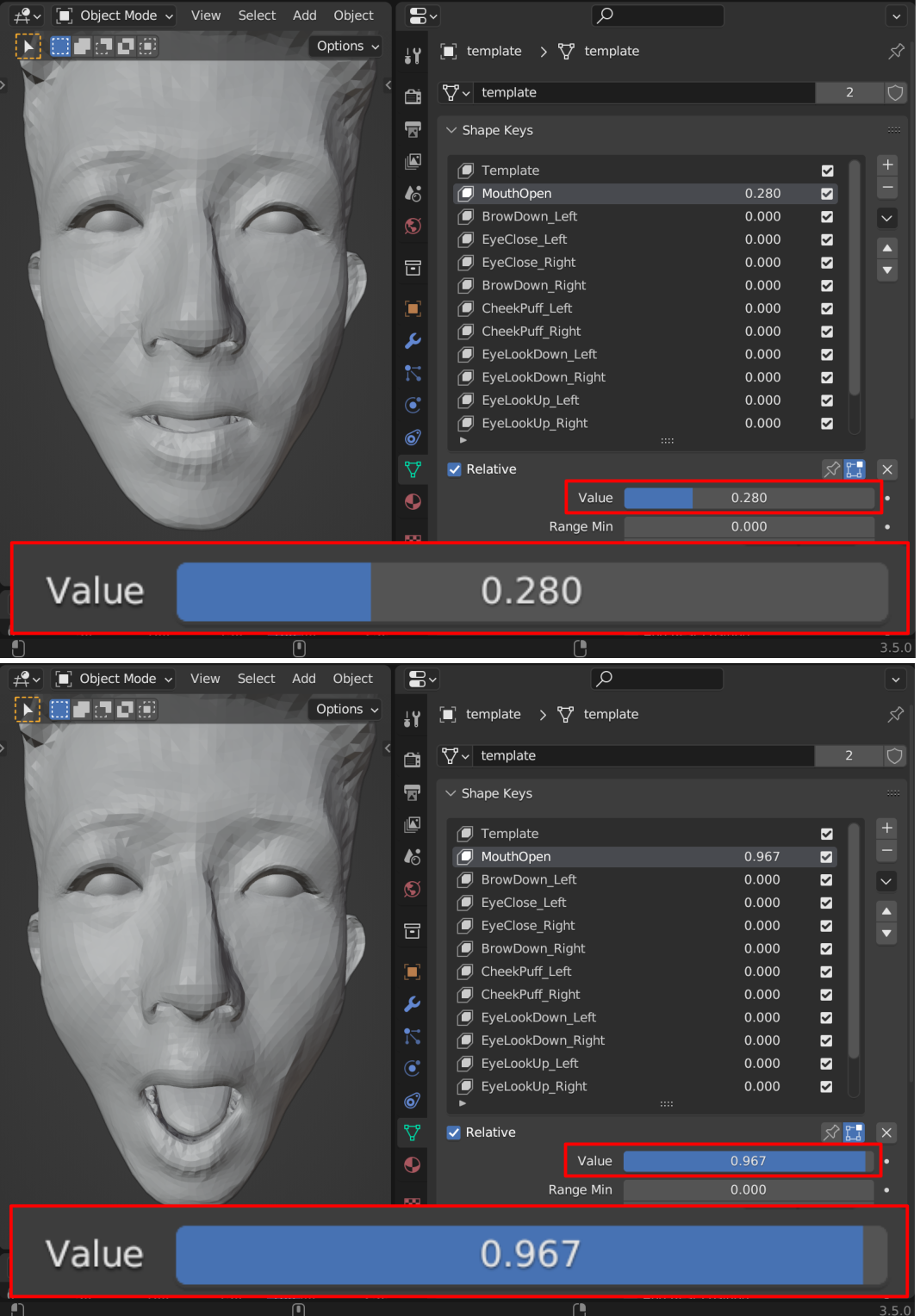}
    \caption{Usage of our blendshapes in Blender.}
    \label{fig:blender}
\end{wrapfigure}
Therefore, it can be directly imported into animation software (such as Blender\cite{Blender}) for synthesizing expressive animations, as depicted in \cref{fig:retargeting}.
We demonstrate the results of the reconstructed facial rig being driven by a performer of a different identity.
During a puckering expression, our method synthesizes distinct lip shapes between individuals (fourth column), and during a grimace, we observe person-specific nasolabial folds (third column).
Expression-related nasolabial folds are properly deactivated when the skin is relaxed (fourth column).
Our facial rig includes complete teeth that can be properly driven (second and fourth column). Due to limited observations, our teeth do not receive vertex deformation. However, constraints on the teeth are considered during optimization to ensure compatibility with lip movements. This ensures that even when updating the expression basis for lip movements, there is no penetration with the teeth.

{\bf Usage in Blender.}
The blendshapes we generate are readily importable into Blender \cite{Blender} for animation as shown in \cref{fig:blender}, where sliders are used for expression adjustments.

\begin{wrapfigure}{r}{0.4\linewidth}
    \centering
    \includegraphics[width=\linewidth]{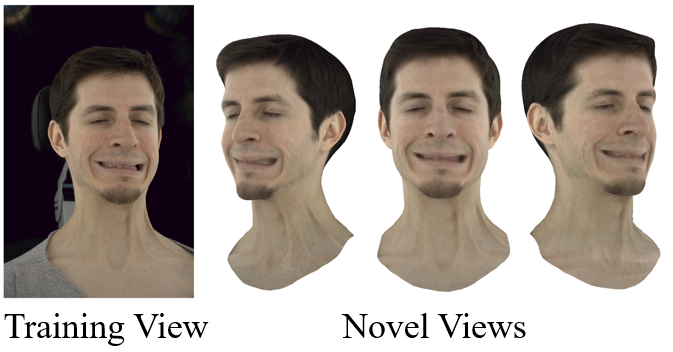}
    \caption{Results of novel view synthesis for an input frame of our method.}
    \label{fig:novel_views}
\end{wrapfigure}

{\bf Novel-view synthesis.}
We demonstrate that our method can synthesize photo-realistic novel views, as shown in \cref{fig:novel_views}.
Our method can accurately reconstruct the geometry and appearance of ears from sparse multi-view inputs, ensuring effective novel-view generalization for ear appearance and synthesizing high-quality ears (third column).
Deferred rendering MLP is suitable for synthesizing photo-realistic facial appearance but cannot be directly imported into current animation software.
Making deferred rendering MLP compatible with animation pipelines is a direction for future work. Recent efforts, such as those presented in \cite{DBLP:conf/cvpr/ChenFHT23,DBLP:journals/tog/ReiserSVSMGBH23,DBLP:journals/corr/abs-2310-17519}, are working towards achieving this goal.
Our method relies on fast mesh-based rasterization and deferred neural rendering, enabling real-time animation and novel-view synthesis.

\section{Limitations}
We aim to reconstruct personalized facial blendshapes from videos, enabling accurate surface geometry.
However, surface geometry is suitable for modeling the skin but not ideal for modeling fine volumetric details like hair.
In the animation pipeline, the focus is primarily on modeling the movement of facial muscles. 
Since hair is not in the region of muscle movement, its impact on animation is relatively small.
However, future work could explore adopting a hybrid representation that uses different geometric forms to express facial skin and hair. This approach could lead to higher-quality rendering of face avatars.
Our method optimizes per-frame head poses, while camera intrinsics and extrinsics are calibrated using a checkerboard pattern once before the capture. Recent works such as \cite{DBLP:conf/cvpr/TruongRMT23,DBLP:journals/corr/abs-2308-10902} hold the potential to integrate with our method to achieve joint estimation of camera parameters.
Our method can personalize template blendshapes. However, the ICT model \cite{ICTli2020learning} used in the experiment does not have a blendshape for the tongue.
Future work could involve testing blendshapes or designing a separate motion approach for the tongue.

\section{Conclusion}

We propose to reconstruct personalized blendshapes from RGB videos via neural inverse rendering, effectively addressing the gap between traditional animation pipelines and cutting-edge neural inverse rendering techniques. 
Leveraging a blendshape rig representation for dynamic facial modeling, we introduce a joint optimization process that refines the rig with per-vertex deformation schemes. This ensures seamless compatibility with animation pipelines and precise alignment with facial performances in RGB videos.
Our contributions extend to an efficient inverse rendering framework that integrates neural shading with blendshapes, enabling the reconstruction of animation-ready facial rigs under diverse lighting and materials. A novel blendshape deformation technique, incorporating differential coordinates augmented with tetrahedral connections and semantic regularization, is introduced to enhance the expressiveness and adherence to volumetric Laplacian regularization.
Experiments showcase the effectiveness of our approach in obtaining high-quality, animation-ready facial rigs from single or sparse multi-view videos, underscoring its accuracy and animation applicability.

\section*{Acknowledgements}
This work was supported by the National Key R\&D Program of China (2023YF\-C3305600), the NSFC (No.62021002), and the Key Research and Development Project of Tibet Autonomous Region (XZ202101ZY0019G). This work was also supported by THUIBCS, Tsinghua University, and BLBCI, Beijing Municipal Education Commission. Feng Xu is the corresponding author.

\bibliographystyle{splncs04}
\bibliography{main}

\begin{thebibliography}{10}
\providecommand{\url}[1]{\texttt{#1}}
\providecommand{\urlprefix}{URL }
\providecommand{\doi}[1]{https://doi.org/#1}

\bibitem{metashape2023}
Agisoft metashape professional (software). \url{http://www.agisoft.com/downloads/installer/} (2023), accessed: 2023-11-16

\bibitem{DBLP:conf/cvpr/AtharXSSS22}
Athar, S., Xu, Z., Sunkavalli, K., Shechtman, E., Shu, Z.: Rignerf: Fully controllable neural 3d portraits. In: {IEEE/CVF} Conference on Computer Vision and Pattern Recognition, {CVPR} 2022, New Orleans, LA, USA, June 18-24, 2022. pp. 20332--20341. {IEEE} (2022). \doi{10.1109/CVPR52688.2022.01972}, \url{https://doi.org/10.1109/CVPR52688.2022.01972}

\bibitem{beeler2011high}
Beeler, T., Hahn, F., Bradley, D., Bickel, B., Beardsley, P.A., Gotsman, C., Sumner, R.W., Gross, M.H.: High-quality passive facial performance capture using anchor frames. ACM Trans. Graph.  \textbf{30}(4), ~75 (2011)

\bibitem{DBLP:journals/corr/abs-2310-17519}
Bharadwaj, S., Zheng, Y., Hilliges, O., Black, M.J., Abrevaya, V.F.: {FLARE:} fast learning of animatable and relightable mesh avatars. CoRR  \textbf{abs/2310.17519} (2023). \doi{10.48550/ARXIV.2310.17519}, \url{https://doi.org/10.48550/arXiv.2310.17519}

\bibitem{DBLP:conf/siggraph/BlanzV99}
Blanz, V., Vetter, T.: A morphable model for the synthesis of 3d faces. In: Waggenspack, W.N. (ed.) Proceedings of the 26th Annual Conference on Computer Graphics and Interactive Techniques, {SIGGRAPH} 1999, Los Angeles, CA, USA, August 8-13, 1999. pp. 187--194. {ACM} (1999), \url{https://dl.acm.org/citation.cfm?id=311556}

\bibitem{DBLP:journals/tog/BouazizWP13}
Bouaziz, S., Wang, Y., Pauly, M.: Online modeling for realtime facial animation. {ACM} Trans. Graph.  \textbf{32}(4),  40:1--40:10 (2013). \doi{10.1145/2461912.2461976}, \url{https://doi.org/10.1145/2461912.2461976}

\bibitem{DBLP:journals/tog/BradleyHPS10}
Bradley, D., Heidrich, W., Popa, T., Sheffer, A.: High resolution passive facial performance capture. {ACM} Trans. Graph.  \textbf{29}(4),  41:1--41:10 (2010). \doi{10.1145/1778765.1778778}, \url{https://doi.org/10.1145/1778765.1778778}

\bibitem{DBLP:conf/nips/CaiFFWZ22}
Cai, H., Feng, W., Feng, X., Wang, Y., Zhang, J.: Neural surface reconstruction of dynamic scenes with monocular {RGB-D} camera. In: NeurIPS (2022), \url{http://papers.nips.cc/paper\_files/paper/2022/hash/06a52a54c8ee03cd86771136bc91eb1f-Abstract-Conference.html}

\bibitem{DBLP:journals/tog/CaoATCSSS21}
Cao, C., Agrawal, V., la~Torre, F.D., Chen, L., Saragih, J.M., Simon, T., Sheikh, Y.: Real-time 3d neural facial animation from binocular video. {ACM} Trans. Graph.  \textbf{40}(4),  87:1--87:17 (2021). \doi{10.1145/3450626.3459806}, \url{https://doi.org/10.1145/3450626.3459806}

\bibitem{DBLP:journals/tog/CaoBZB15}
Cao, C., Bradley, D., Zhou, K., Beeler, T.: Real-time high-fidelity facial performance capture. {ACM} Trans. Graph.  \textbf{34}(4),  46:1--46:9 (2015). \doi{10.1145/2766943}, \url{https://doi.org/10.1145/2766943}

\bibitem{DBLP:journals/tog/CaoHZ14}
Cao, C., Hou, Q., Zhou, K.: Displaced dynamic expression regression for real-time facial tracking and animation. {ACM} Trans. Graph.  \textbf{33}(4),  43:1--43:10 (2014). \doi{10.1145/2601097.2601204}, \url{https://doi.org/10.1145/2601097.2601204}

\bibitem{DBLP:journals/tog/CaoWLZ13}
Cao, C., Weng, Y., Lin, S., Zhou, K.: 3d shape regression for real-time facial animation. {ACM} Trans. Graph.  \textbf{32}(4),  41:1--41:10 (2013). \doi{10.1145/2461912.2462012}, \url{https://doi.org/10.1145/2461912.2462012}

\bibitem{DBLP:conf/svr/CruzT21}
de~Carvalho~Cruz, A.T., Teixeira, J.M.X.N.: A review regarding the 3d facial animation pipeline. In: SVR'21: 23rd Symposium on Virtual and Augmented Reality, Virtual Event, Brazil, October 18 - 21, 2021. pp. 192--196. {ACM} (2021). \doi{10.1145/3488162.3488226}, \url{https://doi.org/10.1145/3488162.3488226}

\bibitem{chaudhuri2020personalized}
Chaudhuri, B., Vesdapunt, N., Shapiro, L., Wang, B.: Personalized face modeling for improved face reconstruction and motion retargeting. In: Computer Vision--ECCV 2020: 16th European Conference, Glasgow, UK, August 23--28, 2020, Proceedings, Part V 16. pp. 142--160. Springer (2020)

\bibitem{DBLP:conf/cvpr/ChenOB023}
Chen, C., O'Toole, M., Bharaj, G., Garrido, P.: Implicit neural head synthesis via controllable local deformation fields. In: {IEEE/CVF} Conference on Computer Vision and Pattern Recognition, {CVPR} 2023, Vancouver, BC, Canada, June 17-24, 2023. pp. 416--426. {IEEE} (2023). \doi{10.1109/CVPR52729.2023.00048}, \url{https://doi.org/10.1109/CVPR52729.2023.00048}

\bibitem{DBLP:conf/cvpr/ChenFHT23}
Chen, Z., Funkhouser, T.A., Hedman, P., Tagliasacchi, A.: Mobilenerf: Exploiting the polygon rasterization pipeline for efficient neural field rendering on mobile architectures. In: {IEEE/CVF} Conference on Computer Vision and Pattern Recognition, {CVPR} 2023, Vancouver, BC, Canada, June 17-24, 2023. pp. 16569--16578. {IEEE} (2023). \doi{10.1109/CVPR52729.2023.01590}, \url{https://doi.org/10.1109/CVPR52729.2023.01590}

\bibitem{Blender}
Community, B.O.: Blender - a 3D modelling and rendering package. Blender Foundation, Stichting Blender Foundation, Amsterdam (2018), \url{http://www.blender.org}

\bibitem{debevec2000acquiring}
Debevec, P., Hawkins, T., Tchou, C., Duiker, H.P., Sarokin, W., Sagar, M.: Acquiring the reflectance field of a human face. In: Proceedings of the 27th annual conference on Computer graphics and interactive techniques. pp. 145--156 (2000)

\bibitem{DBLP:conf/rt/DonnerJ06}
Donner, C., Jensen, H.W.: A spectral {BSSRDF} for shading human skin. In: Akenine{-}M{\"{o}}ller, T., Heidrich, W. (eds.) Proceedings of the Eurographics Symposium on Rendering Techniques, Nicosia, Cyprus, 2006. pp. 409--417. Eurographics Association (2006). \doi{10.2312/EGWR/EGSR06/409-417}, \url{https://doi.org/10.2312/EGWR/EGSR06/409-417}

\bibitem{DBLP:journals/cgf/FyffeNHSBJLD17}
Fyffe, G., Nagano, K., Huynh, L., Saito, S., Busch, J., Jones, A., Li, H., Debevec, P.E.: Multi-view stereo on consistent face topology. Comput. Graph. Forum  \textbf{36}(2),  295--309 (2017). \doi{10.1111/CGF.13127}, \url{https://doi.org/10.1111/cgf.13127}

\bibitem{Gafni_2021_CVPR}
Gafni, G., Thies, J., Zollh{\"o}fer, M., Nie{\ss}ner, M.: Dynamic neural radiance fields for monocular 4d facial avatar reconstruction. In: Proceedings of the IEEE/CVF Conference on Computer Vision and Pattern Recognition (CVPR). pp. 8649--8658 (June 2021)

\bibitem{DBLP:conf/nips/GaoLTRK22}
Gao, H., Li, R., Tulsiani, S., Russell, B., Kanazawa, A.: Monocular dynamic view synthesis: {A} reality check. In: NeurIPS (2022), \url{http://papers.nips.cc/paper\_files/paper/2022/hash/dab5a29f6614ec47ea0ca85c140226fd-Abstract-Conference.html}

\bibitem{DBLP:journals/tog/GaoZXHGZ22}
Gao, X., Zhong, C., Xiang, J., Hong, Y., Guo, Y., Zhang, J.: Reconstructing personalized semantic facial nerf models from monocular video. {ACM} Trans. Graph.  \textbf{41}(6),  200:1--200:12 (2022). \doi{10.1145/3550454.3555501}, \url{https://doi.org/10.1145/3550454.3555501}

\bibitem{DBLP:journals/tog/GarridoVWT13}
Garrido, P., Valgaerts, L., Wu, C., Theobalt, C.: Reconstructing detailed dynamic face geometry from monocular video. {ACM} Trans. Graph.  \textbf{32}(6),  158:1--158:10 (2013). \doi{10.1145/2508363.2508380}, \url{https://doi.org/10.1145/2508363.2508380}

\bibitem{garrido2016reconstruction}
Garrido, P., Zollh{\"o}fer, M., Casas, D., Valgaerts, L., Varanasi, K., P{\'e}rez, P., Theobalt, C.: Reconstruction of personalized 3d face rigs from monocular video. ACM Transactions on Graphics (TOG)  \textbf{35}(3),  1--15 (2016)

\bibitem{ghosh2011multiview}
Ghosh, A., Fyffe, G., Tunwattanapong, B., Busch, J., Yu, X., Debevec, P.: Multiview face capture using polarized spherical gradient illumination. ACM Transactions on Graphics (TOG)  \textbf{30}(6),  1--10 (2011)

\bibitem{grassal2021neural}
Grassal, P., Prinzler, M., Leistner, T., Rother, C., Nie{\ss}ner, M., Thies, J.: Neural head avatars from monocular {RGB} videos. In: {IEEE/CVF} Conference on Computer Vision and Pattern Recognition, {CVPR} 2022, New Orleans, LA, USA, June 18-24, 2022. pp. 18632--18643. {IEEE} (2022). \doi{10.1109/CVPR52688.2022.01810}, \url{https://doi.org/10.1109/CVPR52688.2022.01810}

\bibitem{ichim2015dynamic}
Ichim, A.E., Bouaziz, S., Pauly, M.: Dynamic 3d avatar creation from hand-held video input. ACM Transactions on Graphics (ToG)  \textbf{34}(4),  1--14 (2015)

\bibitem{DBLP:journals/corr/abs-2006-12057}
Kato, H., Beker, D., Morariu, M., Ando, T., Matsuoka, T., Kehl, W., Gaidon, A.: Differentiable rendering: {A} survey. CoRR  \textbf{abs/2006.12057} (2020), \url{https://arxiv.org/abs/2006.12057}

\bibitem{kingma2014adam}
Kingma, D.P., Ba, J.: Adam: A method for stochastic optimization. arXiv preprint arXiv:1412.6980  (2014)

\bibitem{kirschstein2023nersemble}
Kirschstein, T., Qian, S., Giebenhain, S., Walter, T., Nie\ss{}ner, M.: Nersemble: Multi-view radiance field reconstruction of human heads. ACM Trans. Graph.  \textbf{42}(4) (jul 2023). \doi{10.1145/3592455}, \url{https://doi.org/10.1145/3592455}

\bibitem{DBLP:journals/tog/LaineHKSLA20}
Laine, S., Hellsten, J., Karras, T., Seol, Y., Lehtinen, J., Aila, T.: Modular primitives for high-performance differentiable rendering. {ACM} Trans. Graph.  \textbf{39}(6),  194:1--194:14 (2020). \doi{10.1145/3414685.3417861}, \url{https://doi.org/10.1145/3414685.3417861}

\bibitem{lei2023hierarchical}
Lei, B., Ren, J., Feng, M., Cui, M., Xie, X.: A hierarchical representation network for accurate and detailed face reconstruction from in-the-wild images. In: Proceedings of the IEEE/CVF Conference on Computer Vision and Pattern Recognition. pp. 394--403 (2023)

\bibitem{DBLP:conf/eurographics/LewisARZPD14}
Lewis, J.P., Anjyo, K., Rhee, T., Zhang, M., Pighin, F.H., Deng, Z.: Practice and theory of blendshape facial models. In: Lefebvre, S., Spagnuolo, M. (eds.) 35th Annual Conference of the European Association for Computer Graphics, Eurographics 2014 - State of the Art Reports, Strasbourg, France, April 7-11, 2014. pp. 199--218. Eurographics Association (2014). \doi{10.2312/EGST.20141042}, \url{https://doi.org/10.2312/egst.20141042}

\bibitem{li2010example}
Li, H., Weise, T., Pauly, M.: Example-based facial rigging. Acm transactions on graphics (tog)  \textbf{29}(4), ~1--6 (2010)

\bibitem{DBLP:journals/tog/LiYYB13}
Li, H., Yu, J., Ye, Y., Bregler, C.: Realtime facial animation with on-the-fly correctives. {ACM} Trans. Graph.  \textbf{32}(4),  42:1--42:10 (2013). \doi{10.1145/2461912.2462019}, \url{https://doi.org/10.1145/2461912.2462019}

\bibitem{DBLP:journals/tog/LiKZHB020}
Li, J., Kuang, Z., Zhao, Y., He, M., Bladin, K., Li, H.: Dynamic facial asset and rig generation from a single scan. {ACM} Trans. Graph.  \textbf{39}(6),  215:1--215:18 (2020). \doi{10.1145/3414685.3417817}, \url{https://doi.org/10.1145/3414685.3417817}

\bibitem{ICTli2020learning}
Li, R., Bladin, K., Zhao, Y., Chinara, C., Ingraham, O., Xiang, P., Ren, X., Prasad, P., Kishore, B., Xing, J., Li, H.: Learning formation of physically-based face attributes. In: 2020 {IEEE/CVF} Conference on Computer Vision and Pattern Recognition, {CVPR} 2020, Seattle, WA, USA, June 13-19, 2020. pp. 3407--3416. Computer Vision Foundation / {IEEE} (2020). \doi{10.1109/CVPR42600.2020.00347}, \url{https://openaccess.thecvf.com/content\_CVPR\_2020/html/Li\_Learning\_Formation\_of\_Physically-Based\_Face\_Attributes\_CVPR\_2020\_paper.html}

\bibitem{DBLP:conf/iccv/Liu0LL19}
Liu, S., Chen, W., Li, T., Li, H.: Soft rasterizer: {A} differentiable renderer for image-based 3d reasoning. In: 2019 {IEEE/CVF} International Conference on Computer Vision, {ICCV} 2019, Seoul, Korea (South), October 27 - November 2, 2019. pp. 7707--7716. {IEEE} (2019). \doi{10.1109/ICCV.2019.00780}, \url{https://doi.org/10.1109/ICCV.2019.00780}

\bibitem{DBLP:conf/si3d/MaD19}
Ma, L., Deng, Z.: Real-time hierarchical facial performance capture. In: Spencer, S.N., Andrews, S., Tatarchuk, N. (eds.) Proceedings of the {ACM} {SIGGRAPH} Symposium on Interactive 3D Graphics and Games, {I3D} 2019, Montreal, QC, Canada, May 21-23, 2019. pp. 11:1--11:10. {ACM} (2019). \doi{10.1145/3306131.3317016}, \url{https://doi.org/10.1145/3306131.3317016}

\bibitem{mildenhall2020nerf}
Mildenhall, B., Srinivasan, P.P., Tancik, M., Barron, J.T., Ramamoorthi, R., Ng, R.: Nerf: Representing scenes as neural radiance fields for view synthesis. In: ECCV (2020)

\bibitem{mueller2022instant}
M\"uller, T., Evans, A., Schied, C., Keller, A.: Instant neural graphics primitives with a multiresolution hash encoding. ACM Trans. Graph.  \textbf{41}(4),  102:1--102:15 (Jul 2022). \doi{10.1145/3528223.3530127}, \url{https://doi.org/10.1145/3528223.3530127}

\bibitem{Nicolet2021Large}
Nicolet, B., Jacobson, A., Jakob, W.: Large steps in inverse rendering of geometry. ACM Transactions on Graphics (Proceedings of SIGGRAPH Asia)  \textbf{40}(6) (Dec 2021). \doi{10.1145/3478513.3480501}, \url{https://rgl.epfl.ch/publications/Nicolet2021Large}

\bibitem{DBLP:journals/corr/abs-2308-10902}
Park, K., Henzler, P., Mildenhall, B., Barron, J.T., Martin{-}Brualla, R.: Camp: Camera preconditioning for neural radiance fields. CoRR  \textbf{abs/2308.10902} (2023). \doi{10.48550/ARXIV.2308.10902}, \url{https://doi.org/10.48550/arXiv.2308.10902}

\bibitem{semantic_blendshape}
Paul, E., Friesen, W.V.: Facial action coding system: a technique for the measurement of facial movement. Consulting Psychologists  (1978)

\bibitem{qian2024gaussianavatars}
Qian, S., Kirschstein, T., Schoneveld, L., Davoli, D., Giebenhain, S., Nie{\ss}ner, M.: Gaussianavatars: Photorealistic head avatars with rigged 3d gaussians. In: Proceedings of the IEEE/CVF Conference on Computer Vision and Pattern Recognition. pp. 20299--20309 (2024)

\bibitem{DBLP:journals/corr/abs-2007-08501}
Ravi, N., Reizenstein, J., Novotn{\'{y}}, D., Gordon, T., Lo, W., Johnson, J., Gkioxari, G.: Accelerating 3d deep learning with pytorch3d. CoRR  \textbf{abs/2007.08501} (2020), \url{https://arxiv.org/abs/2007.08501}

\bibitem{DBLP:journals/tog/ReiserSVSMGBH23}
Reiser, C., Szeliski, R., Verbin, D., Srinivasan, P.P., Mildenhall, B., Geiger, A., Barron, J.T., Hedman, P.: {MERF:} memory-efficient radiance fields for real-time view synthesis in unbounded scenes. {ACM} Trans. Graph.  \textbf{42}(4),  89:1--89:12 (2023). \doi{10.1145/3592426}, \url{https://doi.org/10.1145/3592426}

\bibitem{DBLP:conf/cvpr/ShaoZTL0L23}
Shao, R., Zheng, Z., Tu, H., Liu, B., Zhang, H., Liu, Y.: Tensor4d: Efficient neural 4d decomposition for high-fidelity dynamic reconstruction and rendering. In: {IEEE/CVF} Conference on Computer Vision and Pattern Recognition, {CVPR} 2023, Vancouver, BC, Canada, June 17-24, 2023. pp. 16632--16642. {IEEE} (2023). \doi{10.1109/CVPR52729.2023.01596}, \url{https://doi.org/10.1109/CVPR52729.2023.01596}

\bibitem{tetgen}
Si, H.: Tetgen, a delaunay-based quality tetrahedral mesh generator. ACM Trans. Math. Softw.  \textbf{41}(2) (feb 2015). \doi{10.1145/2629697}, \url{https://doi.org/10.1145/2629697}

\bibitem{arap}
Sorkine, O., Alexa, M.: As-rigid-as-possible surface modeling. In: Proceedings of the Fifth Eurographics Symposium on Geometry Processing. p. 109–116. SGP '07, Eurographics Association, Goslar, DEU (2007)

\bibitem{DBLP:conf/sgp/SorkineCLARS04}
Sorkine, O., Cohen{-}Or, D., Lipman, Y., Alexa, M., R{\"{o}}ssl, C., Seidel, H.: Laplacian surface editing. In: Boissonnat, J., Alliez, P. (eds.) Second Eurographics Symposium on Geometry Processing, Nice, France, July 8-10, 2004. {ACM} International Conference Proceeding Series, vol.~71, pp. 175--184. Eurographics Association (2004). \doi{10.2312/SGP/SGP04/179-188}, \url{https://doi.org/10.2312/SGP/SGP04/179-188}

\bibitem{sumner2004deformation}
Sumner, R.W., Popovi{\'c}, J.: Deformation transfer for triangle meshes. ACM Transactions on graphics (TOG)  \textbf{23}(3),  399--405 (2004)

\bibitem{DBLP:conf/nips/TancikSMFRSRBN20}
Tancik, M., Srinivasan, P.P., Mildenhall, B., Fridovich{-}Keil, S., Raghavan, N., Singhal, U., Ramamoorthi, R., Barron, J.T., Ng, R.: Fourier features let networks learn high frequency functions in low dimensional domains. In: Larochelle, H., Ranzato, M., Hadsell, R., Balcan, M., Lin, H. (eds.) Advances in Neural Information Processing Systems 33: Annual Conference on Neural Information Processing Systems 2020, NeurIPS 2020, December 6-12, 2020, virtual (2020), \url{https://proceedings.neurips.cc/paper/2020/hash/55053683268957697aa39fba6f231c68-Abstract.html}

\bibitem{tewari2019fml}
Tewari, A., Bernard, F., Garrido, P., Bharaj, G., Elgharib, M., Seidel, H.P., P{\'e}rez, P., Zollhofer, M., Theobalt, C.: Fml: Face model learning from videos. In: Proceedings of the IEEE/CVF Conference on Computer Vision and Pattern Recognition. pp. 10812--10822 (2019)

\bibitem{DBLP:journals/cgf/TewariTMSTWLSML22}
Tewari, A., Thies, J., Mildenhall, B., Srinivasan, P.P., Tretschk, E., Wang, Y., Lassner, C., Sitzmann, V., Martin{-}Brualla, R., Lombardi, S., Simon, T., Theobalt, C., Nie{\ss}ner, M., Barron, J.T., Wetzstein, G., Zollh{\"{o}}fer, M., Golyanik, V.: Advances in neural rendering. Comput. Graph. Forum  \textbf{41}(2),  703--735 (2022). \doi{10.1111/CGF.14507}, \url{https://doi.org/10.1111/cgf.14507}

\bibitem{DBLP:journals/tog/ThiesZN19}
Thies, J., Zollh{\"{o}}fer, M., Nie{\ss}ner, M.: Deferred neural rendering: image synthesis using neural textures. {ACM} Trans. Graph.  \textbf{38}(4),  66:1--66:12 (2019). \doi{10.1145/3306346.3323035}, \url{https://doi.org/10.1145/3306346.3323035}

\bibitem{DBLP:conf/cvpr/ThiesZSTN16}
Thies, J., Zollh{\"{o}}fer, M., Stamminger, M., Theobalt, C., Nie{\ss}ner, M.: Face2face: Real-time face capture and reenactment of {RGB} videos. In: 2016 {IEEE} Conference on Computer Vision and Pattern Recognition, {CVPR} 2016, Las Vegas, NV, USA, June 27-30, 2016. pp. 2387--2395. {IEEE} Computer Society (2016). \doi{10.1109/CVPR.2016.262}, \url{https://doi.org/10.1109/CVPR.2016.262}

\bibitem{DBLP:conf/cvpr/TruongRMT23}
Truong, P., Rakotosaona, M., Manhardt, F., Tombari, F.: {SPARF:} neural radiance fields from sparse and noisy poses. In: {IEEE/CVF} Conference on Computer Vision and Pattern Recognition, {CVPR} 2023, Vancouver, BC, Canada, June 17-24, 2023. pp. 4190--4200. {IEEE} (2023). \doi{10.1109/CVPR52729.2023.00408}, \url{https://doi.org/10.1109/CVPR52729.2023.00408}

\bibitem{DBLP:journals/tog/ValgaertsWBST12}
Valgaerts, L., Wu, C., Bruhn, A., Seidel, H., Theobalt, C.: Lightweight binocular facial performance capture under uncontrolled lighting. {ACM} Trans. Graph.  \textbf{31}(6),  187:1--187:11 (2012). \doi{10.1145/2366145.2366206}, \url{https://doi.org/10.1145/2366145.2366206}

\bibitem{DBLP:journals/mms/VilchisPMG23}
Vilchis, C., P{\'{e}}rez{-}Guerrero, C., Mendez{-}Ruiz, M., Gonz{\'{a}}lez{-}Mendoza, M.: A survey on the pipeline evolution of facial capture and tracking for digital humans. Multim. Syst.  \textbf{29}(4),  1917--1940 (2023). \doi{10.1007/S00530-023-01081-2}, \url{https://doi.org/10.1007/s00530-023-01081-2}

\bibitem{DBLP:journals/tog/VlasicBPP05}
Vlasic, D., Brand, M., Pfister, H., Popovic, J.: Face transfer with multilinear models. {ACM} Trans. Graph.  \textbf{24}(3),  426--433 (2005). \doi{10.1145/1073204.1073209}, \url{https://doi.org/10.1145/1073204.1073209}

\bibitem{wang2020emotion}
Wang, Z., Ling, J., Feng, C., Lu, M., Xu, F.: Emotion-preserving blendshape update with real-time face tracking. IEEE Transactions on Visualization and Computer Graphics  \textbf{28}(6),  2364--2375 (2020)

\bibitem{DBLP:journals/tog/WeiseBLP11}
Weise, T., Bouaziz, S., Li, H., Pauly, M.: Realtime performance-based facial animation. {ACM} Trans. Graph.  \textbf{30}(4), ~77 (2011). \doi{10.1145/2010324.1964972}, \url{https://doi.org/10.1145/2010324.1964972}

\bibitem{DBLP:conf/cvpr/WorchelDHSFE22}
Worchel, M., Diaz, R., Hu, W., Schreer, O., Feldmann, I., Eisert, P.: Multi-view mesh reconstruction with neural deferred shading. In: {IEEE/CVF} Conference on Computer Vision and Pattern Recognition, {CVPR} 2022, New Orleans, LA, USA, June 18-24, 2022. pp. 6177--6187. {IEEE} (2022). \doi{10.1109/CVPR52688.2022.00609}, \url{https://doi.org/10.1109/CVPR52688.2022.00609}

\bibitem{worchel:2022:nds}
Worchel, M., Diaz, R., Hu, W., Schreer, O., Feldmann, I., Eisert, P.: Multi-view mesh reconstruction with neural deferred shading. In: Proceedings of the IEEE/CVF Conference on Computer Vision and Pattern Recognition (CVPR). pp. 6187--6197 (June 2022)

\bibitem{wu2019mvf}
Wu, F., Bao, L., Chen, Y., Ling, Y., Song, Y., Li, S., Ngan, K.N., Liu, W.: Mvf-net: Multi-view 3d face morphable model regression. In: CVPR (2019)

\bibitem{wuu2022multiface}
Wuu, C.h., Zheng, N., Ardisson, S., Bali, R., Belko, D., Brockmeyer, E., Evans, L., Godisart, T., Ha, H., Huang, X., Hypes, A., Koska, T., Krenn, S., Lombardi, S., Luo, X., McPhail, K., Millerschoen, L., Perdoch, M., Pitts, M., Richard, A., Saragih, J., Saragih, J., Shiratori, T., Simon, T., Stewart, M., Trimble, A., Weng, X., Whitewolf, D., Wu, C., Yu, S.I., Sheikh, Y.: Multiface: A dataset for neural face rendering. In: arXiv (2022). \doi{10.48550/ARXIV.2207.11243}, \url{https://arxiv.org/abs/2207.11243}

\bibitem{xiang2024flashavatar}
Xiang, J., Gao, X., Guo, Y., Zhang, J.: Flashavatar: High-fidelity head avatar with efficient gaussian embedding. In: Proceedings of the IEEE/CVF Conference on Computer Vision and Pattern Recognition. pp. 1802--1812 (2024)

\bibitem{DBLP:conf/siggraph/XuWZ0L23}
Xu, Y., Wang, L., Zhao, X., Zhang, H., Liu, Y.: Avatarmav: Fast 3d head avatar reconstruction using motion-aware neural voxels. In: Brunvand, E., Sheffer, A., Wimmer, M. (eds.) {ACM} {SIGGRAPH} 2023 Conference Proceedings, {SIGGRAPH} 2023, Los Angeles, CA, USA, August 6-10, 2023. pp. 47:1--47:10. {ACM} (2023). \doi{10.1145/3588432.3591567}, \url{https://doi.org/10.1145/3588432.3591567}

\bibitem{DBLP:conf/cvpr/Yang0WHSYC20}
Yang, H., Zhu, H., Wang, Y., Huang, M., Shen, Q., Yang, R., Cao, X.: Facescape: {A} large-scale high quality 3d face dataset and detailed riggable 3d face prediction. In: 2020 {IEEE/CVF} Conference on Computer Vision and Pattern Recognition, {CVPR} 2020, Seattle, WA, USA, June 13-19, 2020. pp. 598--607. Computer Vision Foundation / {IEEE} (2020). \doi{10.1109/CVPR42600.2020.00068}, \url{https://openaccess.thecvf.com/content\_CVPR\_2020/html/Yang\_FaceScape\_A\_Large-Scale\_High\_Quality\_3D\_Face\_Dataset\_and\_Detailed\_CVPR\_2020\_paper.html}

\bibitem{zheng2022farl}
Zheng, Y., Yang, H., Zhang, T., Bao, J., Chen, D., Huang, Y., Yuan, L., Chen, D., Zeng, M., Wen, F.: General facial representation learning in a visual-linguistic manner. In: Proceedings of the IEEE/CVF Conference on Computer Vision and Pattern Recognition. pp. 18697--18709 (2022)

\bibitem{zheng2022imavatar}
Zheng, Y., Abrevaya, V.F., Bühler, M.C., Chen, X., Black, M.J., Hilliges, O.: {I} {M} {Avatar}: Implicit morphable head avatars from videos. In: Computer Vision and Pattern Recognition (CVPR) (2022)

\bibitem{Zheng2023pointavatar}
Zheng, Y., Yifan, W., Wetzstein, G., Black, M.J., Hilliges, O.: Pointavatar: Deformable point-based head avatars from videos. In: Proceedings of the IEEE/CVF Conference on Computer Vision and Pattern Recognition (CVPR) (2023)

\bibitem{INSTA:CVPR2023}
Zielonka, W., Bolkart, T., Thies, J.: Instant volumetric head avatars. In: {IEEE/CVF} Conference on Computer Vision and Pattern Recognition, {CVPR} 2023, Vancouver, BC, Canada, June 17-24, 2023. pp. 4574--4584. {IEEE} (2023). \doi{10.1109/CVPR52729.2023.00444}, \url{https://doi.org/10.1109/CVPR52729.2023.00444}

\end{thebibliography}

\title{Supplementary Material for ``High-Quality Mesh Blendshape Generation from Face Videos via Neural Inverse Rendering''} 
\author{}\institute{}

\titlerunning{High-Quality Mesh Blendshape Generation}

\maketitle
\appendix

\section{Discussion of Viewpoint Numbers}
\begin{figure}[tp]
    \centering
    \includegraphics[width=0.7\linewidth]{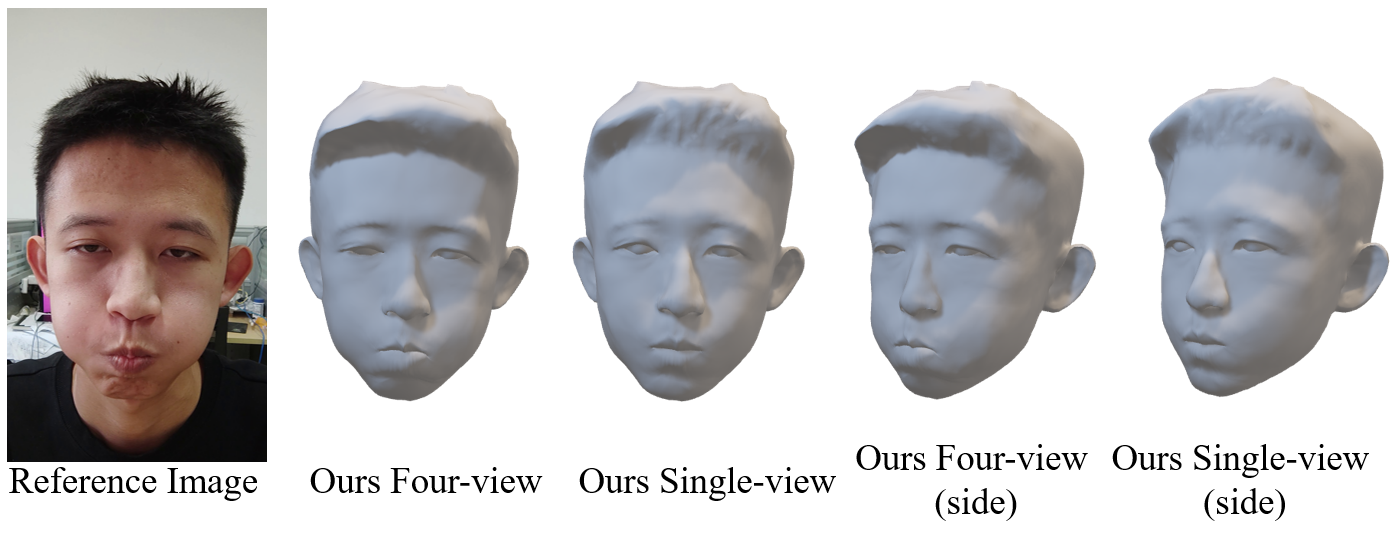}
    \caption{Comparison of geometric reconstruction of the puff expression under single-view and four-view inputs. }%
    \label{fig:4view1view}
\end{figure}

\begin{figure}[tp]
    \centering
    \includegraphics[width=0.8\linewidth]{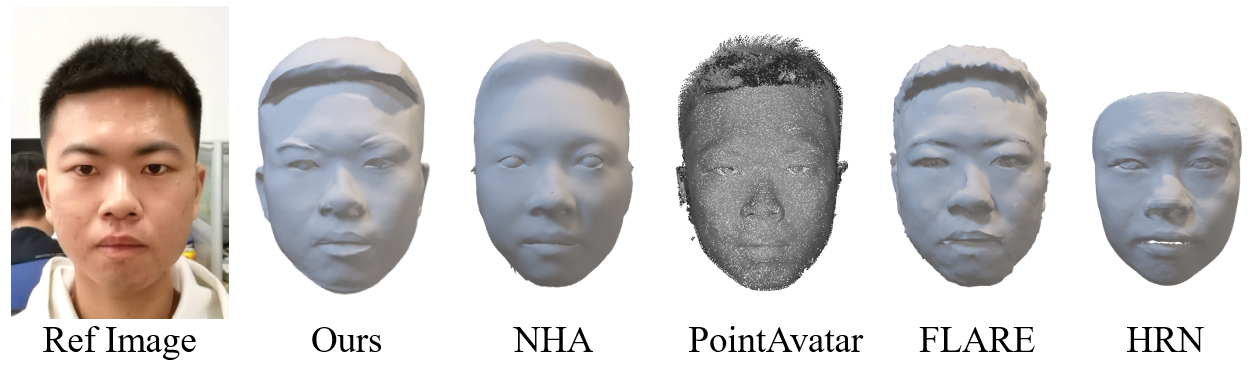}
    \caption{Comparisons of geometry reconstruction between our method and other baselines under single-view input.}
    \label{fig:single_view_compare}
\end{figure}

Our technique gives users the flexibility to use different numbers of cameras to achieve different levels of geometry quality.
As shown in \cref{fig:4view1view}, we qualitatively demonstrate the geometric reconstruction results under single-view and four-view inputs, where under four-view input, we achieve more accurate geometric reconstruction of the cheek-puffing expression compared with the single-view setting.
Additionally, in \cref{table:4viewsvs1view}, we conducted a quantitative evaluation of reconstruction accuracy on the Multiface and NeRSemble datasets.
We found that utilizing four-view inputs further reduces geometric errors compared to single-view inputs, indicating that four views provide a more accurate reconstruction of facial shape. 
\begin{table}[t]
\centering
\begin{tabular}{lllll}
\hline
\multirow{2}{*}{Error(mm)} & \multicolumn{2}{l}{Multiface} & \multicolumn{2}{l}{NeRSemble} \\ \cline{2-5} 
                                          & Mean          & Std           & Mean          & Std           \\ \hline
Single view                                       & 3.02          & 0.16          & 3.58          & 0.37          \\
Four views                                 & \textbf{2.31} & \textbf{0.05} & \textbf{2.73} & \textbf{0.26} \\ \hline
\end{tabular}
\caption{Quantitative comparison of point-to-plane errors between four-view and single-view inputs.}
\label{table:4viewsvs1view}
\end{table}
We choose a four-view setup following \cite{DBLP:conf/cvpr/ShaoZTL0L23} and believe that our method is also applicable to other sparse-view setups.
Please also note that when compared with existing SOTA techniques in \cref{fig:single_view_compare}, our method demonstrates visually superior results in the single-view scenario, indicating that our method also works well with single-view input.

\section{Evaluation of the Neural Regressor}
\begin{figure}[tp]
    \centering
    \includegraphics[width=\linewidth]{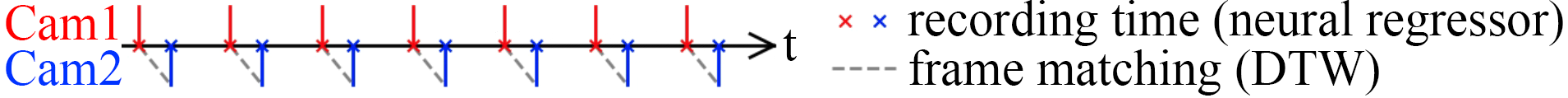}
    \caption{Visualizations of the timeline corresponding to the unsynchronized multi-view inputs.}
    \label{fig:supp_regressor}
\end{figure}

\begin{figure}[tp]
    \centering
    \includegraphics[width=0.6\linewidth]{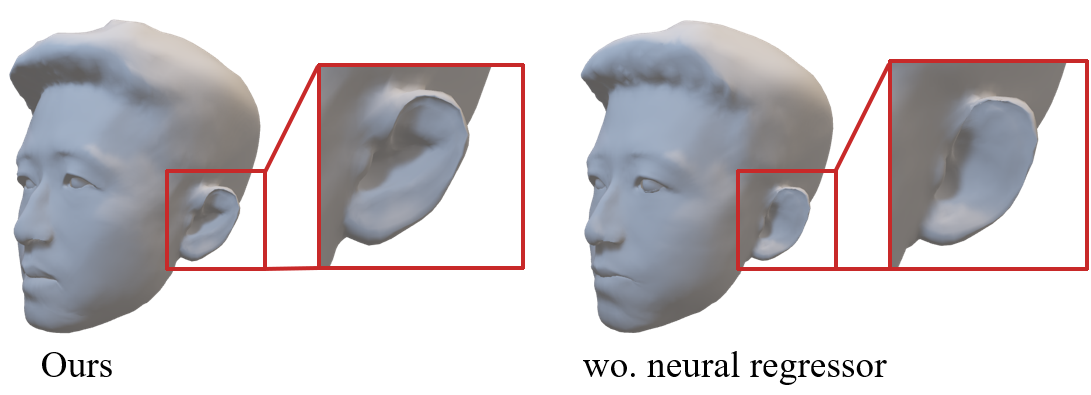}
    \caption{Evaluating the effectiveness of the neural regressor.}
    \label{fig:neural_regressor}
\end{figure}

\begin{figure}[tp]
    \centering
    \includegraphics[width=0.7\linewidth]{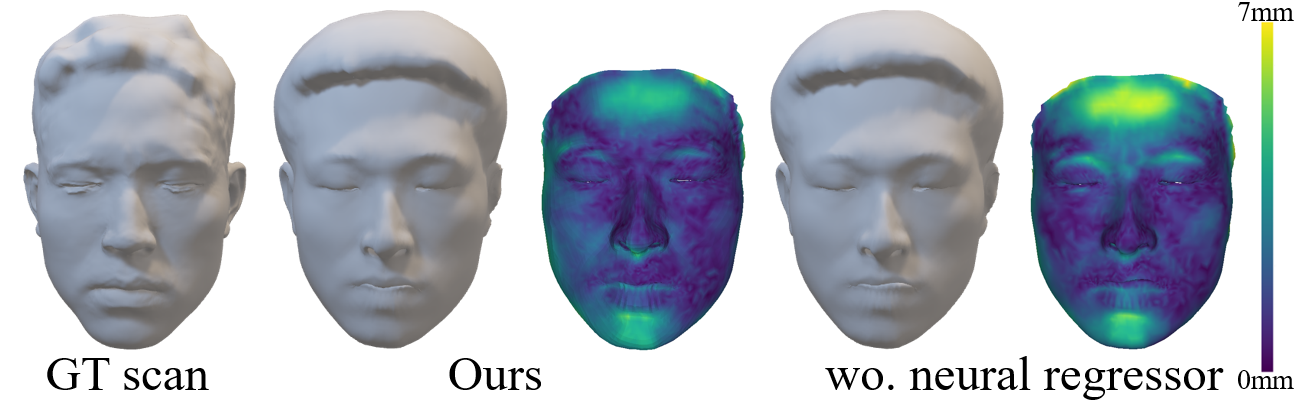}
    \caption{The quantitative reconstruction enhancement obtained by the neural regressor. The second and fourth columns are the reconstruction results, and the third and fifth columns are the point-to-plane error heatmaps compared to the ground truth scan.}
    \label{fig:neural_regressor_quanti}
\end{figure}

In our multi-phone captured data, we cannot precisely synchronize the recording initiation time across devices, so each device records a frame at a slightly different moment. 
As frames from distinct devices are not captured simultaneously, they record varying shapes in motion, necessitating different motion parameters. 
The temporal misalignment can be up to 1/60 second since the frame rate is 30 FPS.
Methods like dynamic time warping could be applied here to match the frames from different views.
However, even if it can always find the temporally closest ones as matches, errors are still there as they are naturally recorded at different times (shown by \cref{fig:supp_regressor}).
The incorporation of the neural regressor allows us to implicitly achieve synchronization across cameras under different views, ensuring temporal accuracy in tracking both expression coefficients and head pose.
Without the regressor, temporal nearest frames 
across devices are erroneously assumed to depict an identical shape, resulting in diminished reconstruction accuracy. 
To assess the impact of the neural regressor, we simply ``synchronize'' videos from different views by finding the temporally nearest frame, and use this as a baseline to compare with our method.
As depicted in \cref{fig:neural_regressor}, our method achieves more accurate geometric details, particularly around the ears.
Furthermore, we quantitatively evaluate the regressor on a subject with ground truth scanned by light stage and observe that including the regressor yields enhanced accuracy, especially in foreheads, as shown in \cref{fig:neural_regressor_quanti}.

\section{Evaluation of Joint Optimizing Blendshapes and Expression Coefficients}
\begin{figure}[tp]
    \centering
    \includegraphics[width=0.6\linewidth]{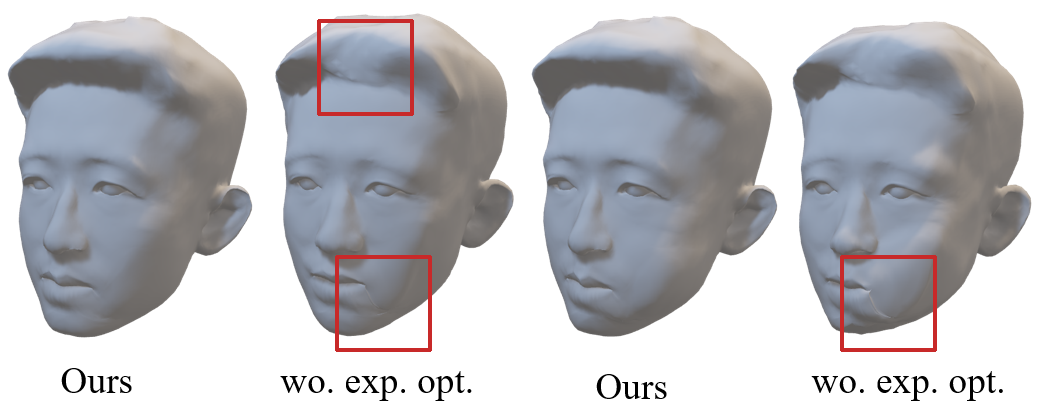}
    \caption{
    Updated blendshapes for neutral and cheek-puffing expressions. The first and third columns show the results of joint optimization, while the second and fourth columns demonstrate the results using a two-stage optimization approach.
    }
    \label{fig:joint_optimization}
\end{figure}

Both blendshapes and expression coefficients affect facial expressions. To address the underdetermined nature of joint optimization, previous works \cite{grassal2021neural, Gafni_2021_CVPR, INSTA:CVPR2023} often adopt a two-stage approach, which stabilizes optimization but does not achieve optimal convergence. In the first stage, a generic expression model is used to fit the expression coefficients \cite{DBLP:conf/cvpr/ThiesZSTN16}. Subsequently, in the second stage, the expression coefficients are fixed and not optimized, focusing solely on optimizing the expression bases.
We propose constraints on the semantic preservation of blendshapes, enabling joint optimization.
To assess the necessity of jointly optimizing expression coefficients and blendshapes, we compare the results obtained by optimizing blendshapes alone with those achieved through joint optimization.
When optimizing blendshapes alone, we derive estimates of the expression coefficients and head poses by fitting the original blendshapes of the ICT face model. Subsequently, we fix the expression coefficients and exclusively optimize the blendshapes.
We illustrate the resulting neutral face and a ``cheek puffing'' blendshape in \cref{fig:joint_optimization}.
In the baseline, artifacts manifest at the corners of the mouth due to inaccurate tracking of the cheek puffing expression during the coefficient estimation stage. 
Additionally, geometric details of the hair are inaccurately embedded into blendshapes.

\section{Baseline Implementation Details}
\label{sec:Baseline Modification}
State-of-the-art facial avatar methods such as Neural Head Avatar \cite{grassal2021neural}, PointAvatar \cite{Zheng2023pointavatar} and FLARE \cite{DBLP:journals/corr/abs-2310-17519} inherently operate with monocular videos as input. To facilitate a fair comparison with our approach, we enhanced these frameworks to accommodate sparse multi-view videos as input.

Specifically, in the data processing phase of these methods, we utilize multi-view videos with calibrated camera parameters as input. During the tracking and optimization stages, we ensure consistency of facial parameters such as shape, poses, and expressions across different viewpoints and optimize these parameters by calculating losses based on images rendered from multiple viewpoints.
After the modification, these methods demonstrated enhanced performance when utilizing multi-view inputs compared to their previous performance with monocular inputs, thus serving as an improved baseline for comparison against our method.

\section{Tetrahedral Connections Establishment}
\label{sec:tetrahedral}
The ICT model features detailed cavities for the mouth, nose, and eyes, with significant facial expressions potentially leading to mesh interpenetration issues. To fully utilize the Laplacian constraint in surface deformation, we establish a connection between the internal cavities and the surface vertices. Specifically, we use tetgen \cite{tetgen} to preprocess the ICT's blendshapes, constructing a small closed volume space between the surface and corresponding internal sockets, where mesh interpenetration is likely to occur. 

\begin{figure}[tp]
    \centering
    \includegraphics[width=0.5\linewidth]{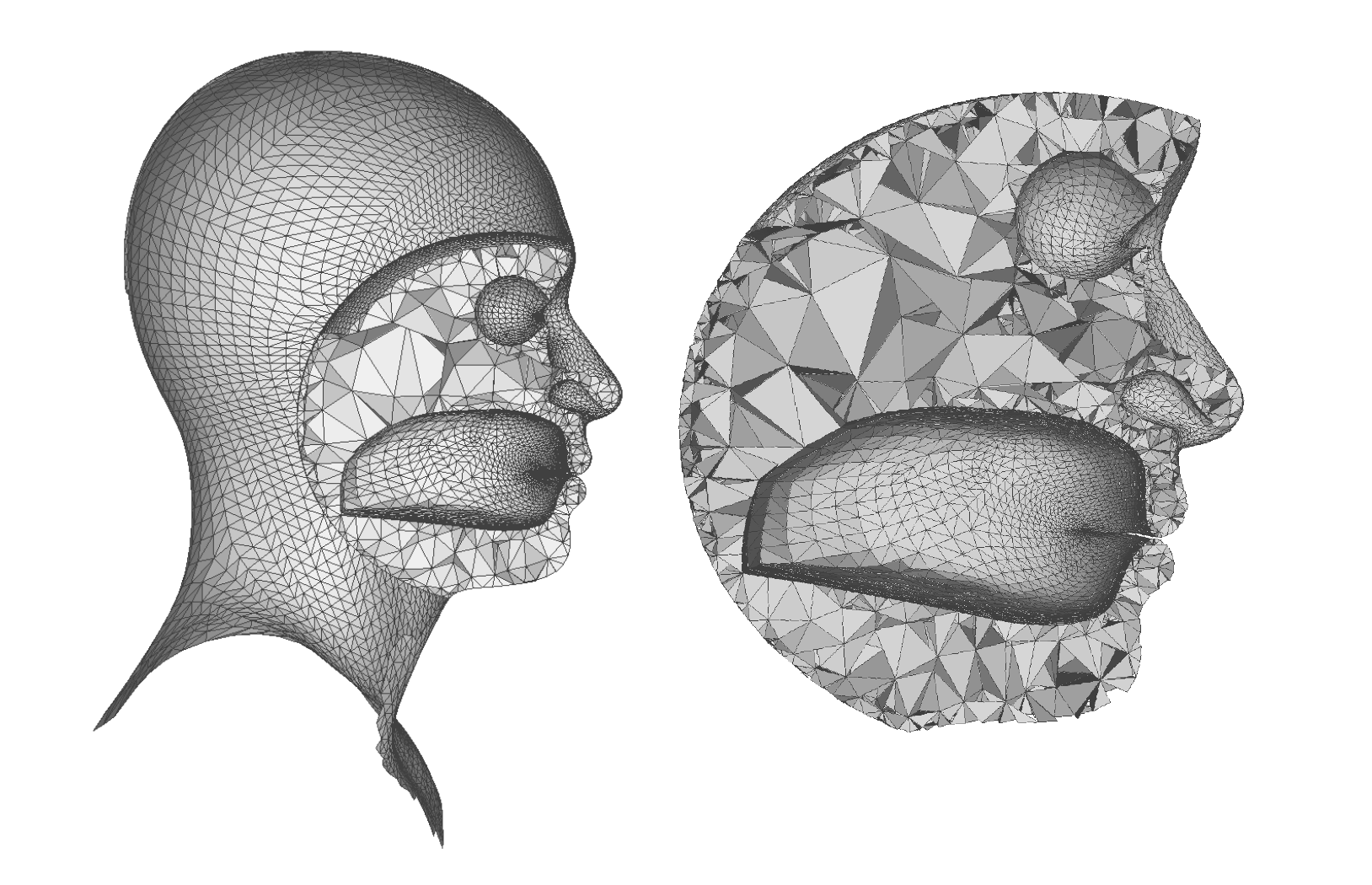}
    \caption{The tetrahedral connections between the cavities and surface.}
    \label{fig:tetgen_detail}
\end{figure}

As illustrated in \cref{fig:tetgen_detail}, we first manually specify the region in the head where we wish to fill tetrahedral connections. 
Tetgen fills this watertight enclosed area with tetrahedrons, whose edges form the connections we intend to establish between the internal cavity and surface vertices. 
Note that even for filling the same enclosed mesh, tetgen's filling results cannot be guaranteed to be topologically consistent each time. Since different blendshapes must share the same topology in our per-vertex deformation technique, 
we first fill the neutral face and then use the ARAP (As-Rigid-As-Possible) \cite{arap} algorithm to deform the tetrahedral vertices to match other blendshapes, creating a set of blendshapes with augmented and consistent topology. 
Moreover, we avoid filling the entire head to reduce computational complexity.

\section{Effective of Differential Coordinates}
\begin{figure}[tp]
    \centering
    \includegraphics[width=0.6\linewidth]{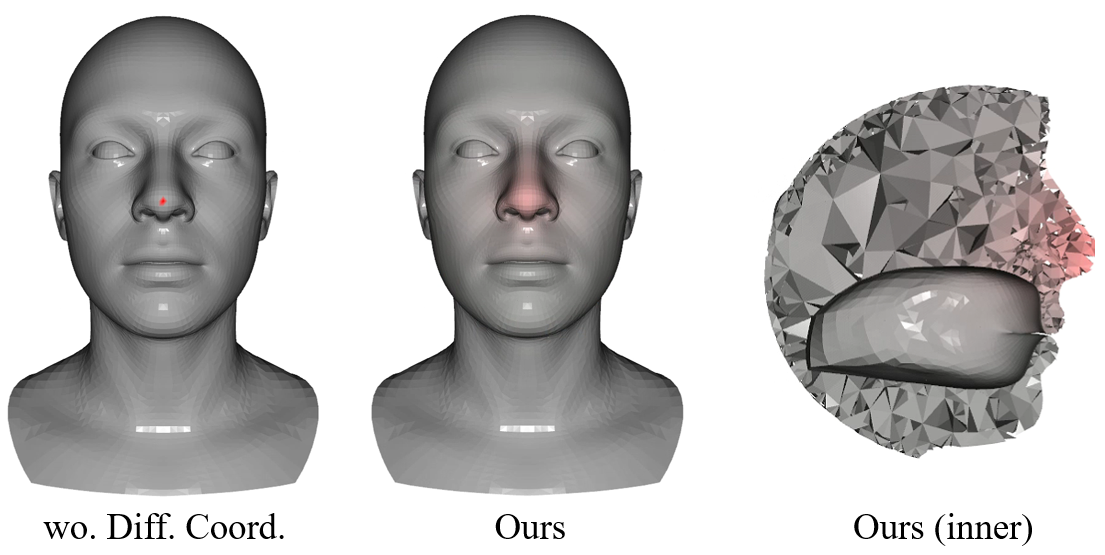}
    \caption{Effectiveness of differential coordinates. The first column shows the gradient obtained at the tip of the nose without using differential coordinates (the gradients are indicated in red). The second column displays the propagated gradients with differential coordinates. The third column presents the gradients for the internal vertices.}
    \label{fig:diff_coord}
\end{figure}

The utilization of differential coordinates propagates vertex gradients to neighboring vertices according to mesh connectivity, thus ensuring the smoothness of per-vertex deformation. As illustrated in \cref{fig:diff_coord}, when the tip of the nose receives a gradient, our method propagates the gradient to the surrounding vertices, including the internally filled vertices. The smoothed gradients facilitate smooth geometry updates while preventing self-intersections.

\section{More Comparisons of Reconstruction Results}
\begin{figure}[tp]
    \centering
    \includegraphics[width=0.8\linewidth]{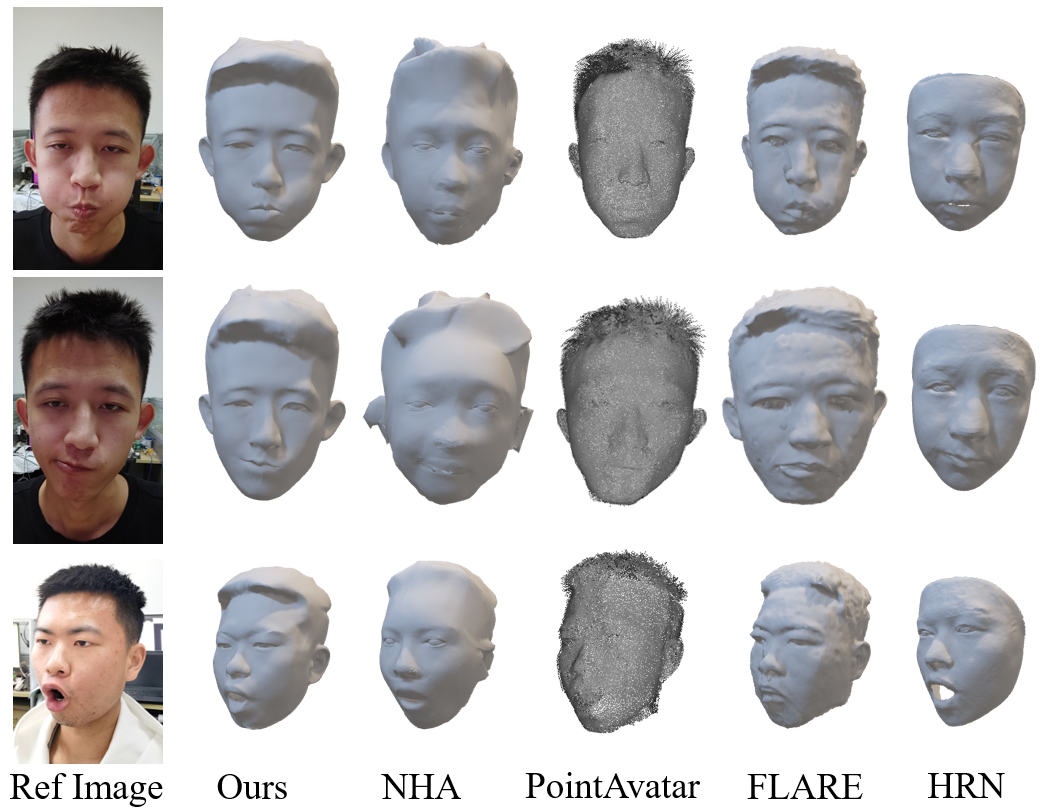}
    \caption{More qualitative comparisons between our method and baselines. All methods accept inputs of four views.}
    \label{fig:more_compare}
\end{figure}

In \cref{fig:more_compare}, we show more reconstruction results to qualitatively compare our method and baselines.
As illustrated in the main text, all methods accept inputs of four views.

\section{More Comparisons of Extracted Blendshapes}
\begin{figure}[tp]
    \centering
    \includegraphics[width=0.8\linewidth]{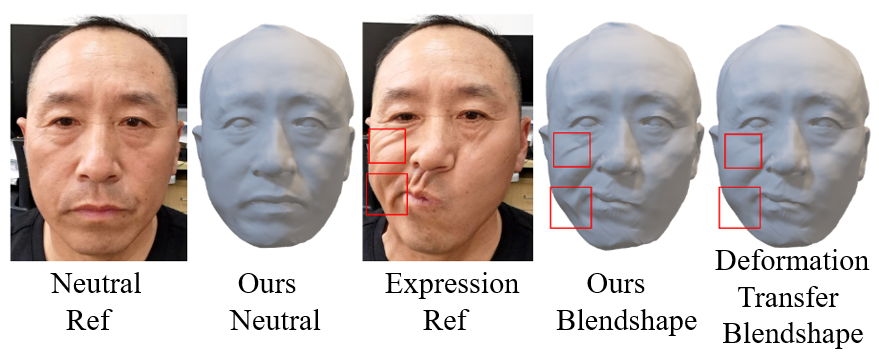}
    \caption{Comparisons of extracted blendshapes with deformation transfer.}
    \label{fig:supp_dt}
\end{figure}

Here, we present a comparison between the blendshapes extracted by our method and deformation transfer \cite{sumner2004deformation} from the neutral. 
As shown in \cref{fig:supp_dt}, our updated blendshape (fourth column) matches the reference image (third column) more closely than deformation transfer (fifth column), which only deforms details in neutral (second column) to the blendshape, lacking expression-specific details such as the wrinkles at the corners of the eyes and mouth (highlighted by red boxes). 

\section{Additional Discussion on Limitations}
\begin{figure}[tp]
    \centering
    \includegraphics[width=0.6\linewidth]{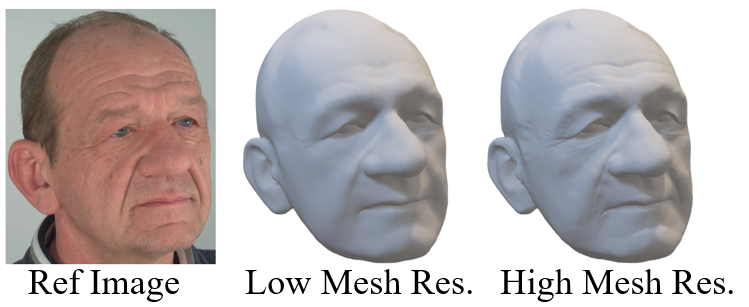}
    \caption{Comparisons of reconstruction results with different mesh resolutions.}
    \label{fig:details}
\end{figure}

\begin{figure}[tp]
    \centering
    \includegraphics[width=0.4\linewidth]{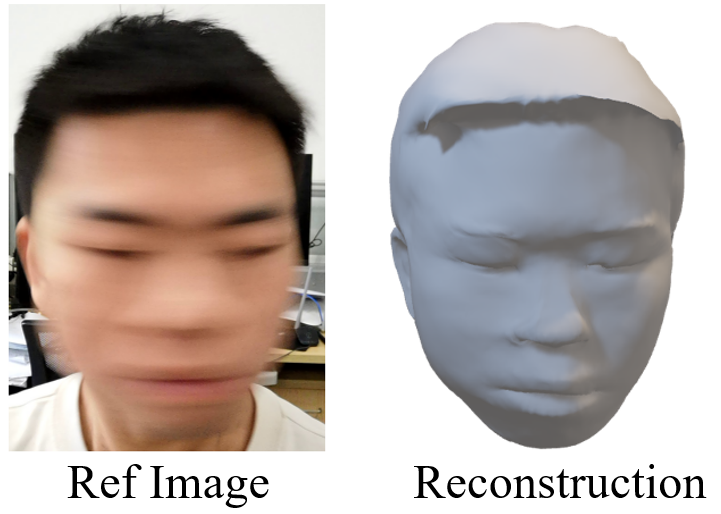}
    \caption{Fast motion leads to quality degradation as blur occurs in the input.}
    \label{fig:fast_motion}
\end{figure}

\begin{figure}[tp]
    \centering
    \includegraphics[width=0.6\linewidth]{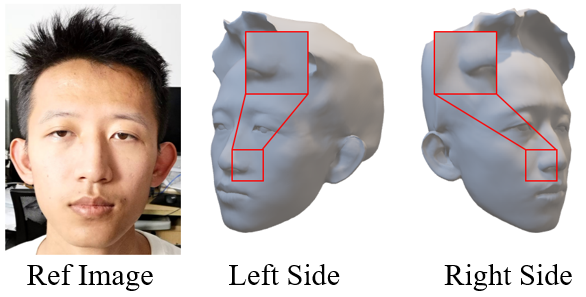}
    \caption{Reconstruction results under strong shadows.}
    \label{fig:harshlight}
\end{figure}

The level of details is constrained by the topology and resolution of the ICT model. Increasing the mesh resolution could help cope this problem, but it comes with increasing training time. As shown in \cref{fig:details}, subdividing the mesh results in better forehead wrinkles and nasolabial folds (second column to third column).
Failure cases may occur when the movement is rapid enough to cause motion blur, as illustrated in \cref{fig:fast_motion}.
Global illumination effects are challenging for our neural radiance methods, such as moving shadows caused by varying head poses. As shown in \cref{fig:harshlight}, the nasal wing in the shadow (second column) is blurrier than the other side (third column).

\section{More Details on Capture Protocols}
The input videos contain around 800 frames with a resolution $504\times896$. 
Our method could support input of any other sizes due to the flexibility of rasterization. 
Subjects are allowed to perform arbitrary expressions, with no need to follow a predefined expression sequence.

\end{document}